\documentclass[letterpaper,twocolumn,10pt]{article}
\usepackage{usenix}

\usepackage{tikz}
\usepackage{amsmath}

\usepackage{filecontents}
\usepackage{nicefrac}
\usepackage{siunitx}
\usepackage{array,framed}
\usepackage{booktabs}
\usepackage{
  color,
  float,
  epsfig,
  wrapfig,
  graphics,
  graphicx,
  subcaption
}
\usepackage{textcomp,amssymb}
\usepackage{setspace}
\usepackage{latexsym,fancyhdr,url}
\usepackage{enumerate}
\usepackage{algorithm2e}
\usepackage{algpseudocode}
\usepackage{graphics}
\usepackage{xparse} % argument parsing -- \edist
\usepackage{xspace}
\usepackage{multirow}
\usepackage{csvsimple}
\usepackage{balance}
\usepackage[]{todonotes}
\usepackage{booktabs} 
\usepackage{pifont}
\usepackage{tabularx}
\usepackage{multicol}
\usepackage{xurl}
\usepackage[most]{tcolorbox}
\usepackage{
  tikz,
  pgfplots,
  pgfplotstable
}
\usepackage{hyperref}
\usepackage{makecell,pifont,array,ragged2e}

\usepackage{
  tikz,
  pgfplots,
  pgfplotstable
}
\usepackage{pgfplots}

\usetikzlibrary{
  shapes.geometric,
  arrows,
  external,
  pgfplots.groupplots,
  matrix
}
\usepackage{mathtools}

\pgfplotsset{compat=1.9}
\usepackage{tikz}
\usepackage{xcolor}
\usepackage{amssymb}        % \checkmark
\usetikzlibrary{
  shapes.geometric,
  shapes.misc,
  shapes.symbols,
  arrows.meta,
  positioning,
  calc,
  fit,
  patterns,
  decorations.pathmorphing
}
 
\definecolor{att}{RGB}{198,55,55}
\definecolor{sys}{RGB}{70,80,100}
\definecolor{ok}{RGB}{50,150,80}
\definecolor{bg}{RGB}{248,248,250}
 
\tikzset{
  >={Stealth[length=2.6mm,width=2mm]},
  font=\sffamily\small,
  attbox/.style ={draw=att, fill=att!8,  thick, rounded corners=2pt},
  sysbox/.style ={draw=sys, fill=sys!5,  thick, rounded corners=3pt},
  okbox/.style  ={draw=ok,  fill=ok!8,   thick, rounded corners=2pt},
  flow/.style   ={->, thick, draw=sys},
  attflow/.style={->, thick, draw=att, dashed},
  okflow/.style ={->, thick, draw=ok},
  tag/.style    ={font=\sffamily\bfseries},
  caption/.style={font=\sffamily\bfseries\small, align=center},
}
 
\newcommand{\tinyface}[2]{%
  \begin{scope}[shift={#1}]
    \draw[draw=#2,thick] (0,0) circle (0.18);
    \fill[#2] (-0.07,0.05) circle (0.02);
    \fill[#2] ( 0.07,0.05) circle (0.02);
    \draw[#2] (-0.07,-0.04) .. controls (0,-0.09) .. (0.07,-0.04);
  \end{scope}}
 
\newcommand{\phoneframe}[3]{%
  \draw[attbox] ($#1+(-#2,-#3)$) rectangle ($#1+(#2,#3)$);
  \draw[fill=white, draw=att!40]
        ($#1+(-#2+0.1,-#3+0.12)$) rectangle ($#1+(#2-0.1,#3-0.12)$);}
 
\newcommand{\camera}[2]{%
  \begin{scope}[shift={#1}, scale=#2, every path/.style={thick}]
    \draw[sys, fill=sys!15] (-0.18,-0.55) rectangle (0.18,-0.36);
    \draw[sys, fill=sys!10]
        (-0.42,-0.70) -- (0.42,-0.70) -- (0.32,-0.55) -- (-0.32,-0.55) -- cycle;
    \draw[sys, fill=sys!10]
        (-0.5,-0.36) -- (0.5,-0.36) arc (-90:90:0.36)
        -- (-0.5,0.36) arc (90:270:0.36);
    \draw[sys, fill=white]  (0,0) circle (0.30);
    \draw[sys, fill=sys!30] (0,0) circle (0.22);
    \draw[sys, fill=sys!55] (0,0) circle (0.12);
    \fill[white] (-0.08,0.08) circle (0.05);
    \fill[ok] (0.65,0.15) circle (0.045);
  \end{scope}}
 
\newcommand{\browser}[3]{%
  \begin{scope}[shift={#1}]
    \draw[sysbox] (-#2,-#3) rectangle (#2,#3);
    \draw[fill=sys!12, draw=sys] (-#2,#3-0.25) rectangle (#2,#3);
    \fill[att]    (-#2+0.15,#3-0.125) circle (0.05);
    \fill[orange] (-#2+0.30,#3-0.125) circle (0.05);
    \fill[ok]     (-#2+0.45,#3-0.125) circle (0.05);
    \draw[fill=white, draw=sys!50]
       (-#2+0.15,#3-0.55) rectangle (#2-0.15,#3-0.35);
    \draw[sys!70] (-#2+0.15,#3-0.80) -- (#2-0.15,#3-0.80);
    \draw[sys!70] (-#2+0.15,#3-1.05) -- (#2*0.4,#3-1.05);
  \end{scope}}
 
\newcommand{\laptop}[2]{%
  \begin{scope}[shift={#1}, scale=#2, every path/.style={thick}]
    \draw[attbox] (-0.45,0.02) rectangle (0.45,0.62);
    \draw[fill=att!15, draw=att!50] (-0.38,0.08) rectangle (0.38,0.56);
    \draw[attbox, fill=att!12]
        (-0.62,-0.08) -- (0.62,-0.08) -- (0.5,0.02) -- (-0.5,0.02) -- cycle;
  \end{scope}}
 
\newcommand{\tokenkey}[2]{%
  \begin{scope}[shift={#1}, scale=#2, every path/.style={thick}]
    \draw[ok, fill=ok!25] (0,0) circle (0.22);
    \draw[ok, fill=white] (0,0) circle (0.08);
    \draw[ok, fill=ok!25]
        (0.18,-0.07) -- (0.95,-0.07) -- (0.95,0.07) -- (0.18,0.07) -- cycle;
    \draw[ok, fill=ok!25]
        (0.60,-0.07) -- (0.60,-0.22) -- (0.68,-0.22) -- (0.68,-0.07) -- cycle;
    \draw[ok, fill=ok!25]
        (0.80,-0.07) -- (0.80,-0.26) -- (0.88,-0.26) -- (0.88,-0.07) -- cycle;
  \end{scope}}
 
\newcommand{\cookieicon}[3]{%
  \begin{scope}[shift={#1}, scale=#2]
    \draw[fill=#3!25, draw=#3, thick] (0,0) circle (0.24);
    \fill[#3] (-0.10,0.07) circle (0.030);
    \fill[#3] ( 0.09,-0.06) circle (0.030);
    \fill[#3] ( 0.05,0.12) circle (0.026);
    \fill[#3] (-0.11,-0.10) circle (0.028);
    \fill[#3] ( 0.13,0.09) circle (0.022);
  \end{scope}}
 
\usepackage{mathtools}

\DeclareMathAlphabet{\mathcal}{OMS}{cmsy}{m}{n}
\DeclareGraphicsExtensions{%
    .png,.PNG,%
    .pdf,.PDF,%
    .jpg,.mps,.jpeg,.jbig2,.jb2,.JPG,.JPEG,.JBIG2,.JB2}

\usepackage{makecell}

\newcommand{\mypar}[1]{\smallskip\noindent\textbf{#1.}\xspace}
\newcommand{\mypara}[1]{\smallskip\noindent\textit{#1.}\xspace}

\newcommand{\ageverif}{\textit{VP1}\xspace}
\newcommand{\agego}{\textit{VP2}\xspace}
\newcommand{\gotaca}{\textit{VP3}\xspace}
\newcommand{\yoti}{\textit{VP4}\xspace}
\newcommand{\verifymy}{\textit{VP5}\xspace}
\newcommand{\friendlyid}{\textit{VP6}\xspace}
\newcommand{\oneid}{\textit{VP7}\xspace}
\newcommand{\paymentico}{\textit{VP8}\xspace}
\newcommand{\anonymage}{\textit{VP9}\xspace}
\newcommand{\veratad}{\textit{VP10}\xspace}
\newcommand{\pleenk}{\textit{VP11}\xspace}
\newcommand{\gocam}{\textit{VP12}\xspace}
\newcommand{\ondato}{\textit{VP13}\xspace}
\newcommand{\agerify}{\textit{VP14}\xspace}
\newcommand{\verifyage}{\textit{VP15}\xspace}

\newcommand{\cmarks}{\textcolor{green!60!black}{\checkmark}} % Checkmark
\newcommand{\xmarks}{\textcolor{red!60!black}{$\times$}}     % Cross
\newcommand{\wmarks}{\textcolor{orange}{$\sim$}}              % Partial/Weak
\newcommand{\na}{\textemdash}
\newcolumntype{Y}{>{\RaggedRight\arraybackslash}X} % wraps + allows hyphenation

\newtcolorbox[auto counter, number within=section]{answer}{%
  colback=gray!10, % Background color
  colframe=black, % Border color
  toprule=-0.25mm, % Remove top border
  bottomrule=-0.25mm, % Remove bottom border
  leftrule=1.5mm, % Add left border
  rightrule=-0.25mm, % Remove right border
  left=0.1mm, right=0.1mm, top=0.1mm, bottom=0.1mm, % Padding
  enhanced,
  arc=0mm,
  breakable = true,
}

\begin{document}
%-------------------------------------------------------------------------------

%don't want date printed
\date{}

\title{\Large \bf X-rated Compliance Theater: An Empirical Evaluation of European Age Verification Systems in Adult Websites}

%for single author (just remove % characters)
\author{
Simone Lavermicocca \and
Michele Carminati \and
Stefano  Longari \\
Dipartimento di Elettronica, Informazione e Bioingegneria \\
Politecnico di Milano, Milan, Italy
} % end author

\maketitle

\begin{abstract}
Age verification is rapidly emerging as a central regulatory instrument for protecting minors online, with several jurisdictions mandating its deployment for access to adult and pornographic content. This regulatory direction raises significant privacy concerns, as it risks binding sensitive content access to identity-related attributes. It also introduces security risks, since age-verification mechanisms are often outsourced to third-party providers with limited transparency into the robustness of their verification processes.
In this work, we conduct, to the best of our knowledge, the first exploratory security assessment of regulation-mandated age-verification mechanisms deployed by adult websites. Rather than treating age verification as a purely regulatory question, we empirically examine whether current deployments provide security guarantees commensurate with the privacy risks of relying on sensitive identity-related data. Our methodology combines ecosystem mapping, adversary modeling, and empirical testing across four countries, covering document-based verification, biometric age estimation, indirect signals, and website-workflow integration.
Our results reveal systemic weaknesses across mechanisms and integrations under realistic threat assumptions, including failures against low-cost, widely accessible attacks. Finally, we derive concrete guidelines and design directions for mitigating the security and privacy risks exposed by current age-verification deployments.

\end{abstract}

\section{Introduction}
\label{sec:intro}

Governments worldwide, and particularly in Europe, are rapidly mandating online age verification for access to adult content. At the EU level, instruments such as the Digital Services Act have strengthened duties to protect minors online and accelerated expectations around age assurance for age-restricted material~\cite{eu_avmsd_2018,eu_dsa_2022}. National child-protection regimes in France, Germany, Italy, and the UK have introduced more explicit age-gating requirements and enforcement powers for online pornography~\cite{arcom_referentiel_2024,kjm_avs_raster_2022,uk_online_safety_act_2023,oecd_age_assurance_landscape_2025}. The regulatory landscape and industry adoption are still evolving, with major adult websites changing, delaying, or inconsistently deploying age-verification mechanisms as enforcement pressure develops.

Although age verification is increasingly discussed beyond pornography, including social media, gaming, communication platforms, and restricted media content, pornographic websites are a particularly critical domain because failures in this context expose an especially sensitive link between identity-related attributes and access to sexually explicit material.

Enforcement pressure has driven many adult websites to integrate a heterogeneous ecosystem of third-party age-assurance providers, whose mechanisms range from live facial age estimation to identification documents or credit-card checks. This regulatory trajectory raises significant security and privacy concerns. Age-verification workflows combine highly sensitive content consumption with identity attributes such as government IDs, biometrics, or financial metadata. Such coupling creates an unusually critical asset: a breach or misuse would expose individuals to risks beyond those in less sensitive contexts. Large data breaches in adjacent ecosystems illustrate this impact~\cite{eff-au10tix-2024}, while most providers operate without audited guarantees about data minimization, retention, or linkage~\cite{unicef_day_2021_ageassurance,aepd_decalogue_2023,edri_age_verification_2023}. At the same time, the practical security posture of deployed mechanisms remains rarely analyzed, and many rely on pipelines whose robustness to adversarial tampering is unknown or fragile when scrutinized~\cite{tschirsich_videoident_2022,sharma_selwal_pad_2023}.

In this paper, we investigate whether age-verification systems deployed on adult websites provide effective security guarantees. Although a critique of this reasoning may be that regulators aim to reduce incidental or low-effort access rather than defeat determined adversary, we argue that robustness nonetheless matters because deployed systems impose real privacy costs on adult users and create valuable flows of sensitive identity-related data. If they can be bypassed with little effort, their privacy burden may not be justified by corresponding protective benefits. Our adversarial model does not imply that all policy goals require resistance to all determined attackers; rather, it tests the stronger security expectations implied when age-verification systems are deployed at population scale, presented as protective mechanisms, and used to justify processing sensitive identity-related information.

To achieve this goal, we present the first systematic empirical study of how age verification is implemented in practice on European adult websites. Our study surveys deployments across 154 adult websites in four countries, derives the corresponding adversary models, and evaluates whether, how, and by whom they can be circumvented. We study two attacker strategies: (i) \emph{mechanism-level} attacks that generate a valid proof without completing legitimate verification, and (ii) \emph{deployment-level} bypasses that avoid verification by reusing someone else's proof/token or exploiting flaws in website integration.

Taken together, our evidence indicates that many current age-verification deployments operate closer to compliance-oriented controls than to robust protective mechanisms, while still imposing substantial privacy risks. In summary, we contribute by:
\begin{itemize}
\item Characterizing the European age-verification ecosystem by mapping key providers, deployed mechanisms, and relevant adversaries.
\item Conducting the first systematic exploratory security assessment of regulation-mandated age-verification systems for adult websites across four European countries.
\item Providing guidelines and design directions to help future deployments align with realistic adversary models and avoid introducing privacy risks without corresponding protective benefits.
\end{itemize}    % basic introduction
\section{Background}
\label{sec:background}

\mypar{Regulatory Context}
Recent years have seen a global acceleration in laws and regulatory initiatives requiring online services to mitigate minors' access to pornography and other age-restricted content. Comparative policy surveys document a heterogeneous legal landscape across Europe, the UK, and other jurisdictions, reflecting both child-protection objectives and increasing willingness to impose concrete compliance duties on content and platform operators~\cite{oecd_age_assurance_landscape_2025,eu_dsa_2022}. In the EU, the Digital Services Act (DSA) establishes horizontal duties to protect minors and manage systemic risks, while national authorities adopt more specific instruments for adult content~\cite {eu_dsa_2022,ecMinorsGuidelines2025}. In the UK, public guidance accompanying the Online Safety Act emphasizes both age checks and measures against content promoting workarounds, such as VPN-based circumvention aimed at young users~\cite{uk_gov_osa_explained_2025}.

As shown in Table~\ref{tab:regulatory-methods}, regulatory regimes differ in whether they prescribe specific methods and which methods they deem acceptable or effective. Some frameworks remain method-neutral, specifying outcomes such as preventing minors' access, while others define “highly effective” approaches or impose architectural constraints~\cite{oecd_age_assurance_landscape_2025,euAgeAssuranceMapping2024}. Consequently, allowed mechanisms vary across jurisdictions and commonly include government-ID checks, document-plus-selfie proofing, payment-instrument checks, third-party attribute attestations, and biometric age estimation~\cite{oecd_age_assurance_landscape_2025,euAgeAssuranceMapping2024}. Some regimes require multiple compliant options to cover different user populations, while others accept narrower method sets if they satisfy an effectiveness threshold~\cite{oecd_age_assurance_landscape_2025,arcomBenchmark2024}.

Privacy is acknowledged as a constraint, but usually through general data-protection principles such as proportionality and minimization, rather than through mandatory privacy-preserving architectures~\cite{eu_dsa_2022,ecMinorsGuidelines2025}. Among the instruments surveyed here, only the French technical benchmark and the Italian non-technical guidelines explicitly require at least one “double anonymity” solution, designed to prevent any single actor from linking civil identity with pornographic-content access~\cite{arcomBenchmark2024,agcom_age_verification_2025}. Other authorities frame anonymity and unlinkability as design goals or best practices, without making “double anonymity” a formal compliance requirement~\cite{aepd_decalogue_2023,ecMinorsGuidelines2025}.

\begin{table*}[tb]
\centering
\footnotesize
\setlength{\tabcolsep}{3pt}
\newcommand{\cmark}{\ding{51}}
\newcommand{\xmark}{\ding{55}}
\caption{
Comparison of European regulatory instruments and age-verification approaches. “Reusability” means tying checks to persistent accounts or reusable tokens. “Website verification” indicates if adult sites may perform first-party age verification.
}

\resizebox{\linewidth}{!}{
\begin{tabular}{lccccccccccc}
\toprule
\multirow{2}{*}{\textbf{Instrument}} &
\textbf{Technical} &
\multirow{2}{*}{\textbf{eID}} &
\textbf{Document} &
\textbf{Age} &
\multirow{2}{*}{\textbf{Phone}} &
\multirow{2}{*}{\textbf{Email}} &
\textbf{Credit} &
\textbf{In person} &
\multirow{2}{*}{\textbf{Reusability}} &
\multirow{2}{*}{\textbf{Website}} & 
\textbf{Double} \\
&
\textbf{guidance} &
 &
\textbf{upload} &
\textbf{estimation} &
 &
 &
\textbf{Card} &
\textbf{verification} &
 &
 &
\textbf{anonymity} \\
\midrule
KJM AVS-Raster (DE, 2022)~\cite{kjm_avs_raster_2022} &
yes &
\cmark &
\cmark & 
\cmark &  
&  
&  
&  
\cmark & 
\cmark &
\cmark & 
\\

ARCOM reference (FR, 2024)~\cite{arcom_referentiel_2024} &
yes &
\cmark &
\cmark &
\cmark &
 &
 &
\cmark &
\cmark &
\cmark* &
 &
\cmark \\

Ofcom HEAA guidance (UK, 2025)~\cite{ofcom_heaa_2025} &
yes & 
\cmark &
\cmark &
\cmark &
\cmark &
\cmark &
\cmark &
 &
\cmark &
\cmark &
  \\

EU DSA Art.~28 Guidelines (EU, 2025)~\cite{ecMinorsGuidelines2025} &
Guidelines only &
\cmark &
\cmark &
\cmark &
 &
 &
 &
 &
\cmark &
** &
\cmark  \\
AGCOM (IT)~\cite{agcom_age_verification_2025} & 
no &
&
 &
 &
 &
 &
 &
 &
& 
& \cmark
 \\
\bottomrule
\multicolumn{12}{l}{* the websites are explicitly forbidden to bind verification to the website account; ** The EU guidelines encourage but do not mandate third party verification.} 
\end{tabular}
}
\label{tab:regulatory-methods}
\end{table*}

\mypar{Age Verification Mechanisms Taxonomy} 
Based on regulations, a preliminary analysis, and prior work, we identify five approaches for age verification on adult websites.

\mypara{Digital Identity and eID-Based Verification}
Digital identity approaches are frequently cited as a high-assurance model for age verification. These mechanisms rely on national eID frameworks, most notably those defined under eIDAS and the forthcoming EU Digital Identity Wallet, to attest age attributes derived from government-issued credentials. In principle, such systems offer strong security guarantees through prior identity verification and robust fraud controls. However, their direct applicability to adult content access remains limited, as existing deployments were not designed for privacy-preserving, unlinkable age assertions and risk coupling sensitive content access with real-world identities.

\mypara{Age Estimation} AI-based Age estimation systems require the user to present their face to a camera, allowing the system to infer an approximate age from facial features. Their popularity stems from their perceived ease of use and minimal user friction, but they also raise questions about accuracy, demographic bias, and the handling of biometric data. They have also become a visible target for practical circumvention: recent public reports describe users bypassing UK age-verification checks on platforms such as Reddit and Discord by presenting videogame photo-mode avatars to facial age-estimation systems~\cite{verge_deathstranding_ageverification_2025}.

\mypara{ID-Based Verification} ID-based mechanisms require users to submit an identification document, typically chosen from a country-specific list of valid documents. Submission may occur through a live camera capture or by uploading a static picture. In some implementations, they are paired with a selfie, intended to verify document ownership.

\mypara{Email-, SIM-, and Payment-Based Evidence}
Some regulations permit the use of indirect signals, such as the age of an email account, sending a confirmation link to verify ownership, and using the account’s creation date as a proxy for the user’s age. Some countries also permit mechanisms based on financial or telecom ownership, e.g., performing a micro-transaction on a credit card or sending an SMS to a phone number, under the assumption that possession of these resources typically correlates with being of legal age.

\mypara{Physical Tickets}
Users can buy an in-person ticket after an ID check by shop staff and redeem it online. This avoids transmitting ID documents digitally, but requires physical presence and relies on merchant compliance.

\mypar{Privacy Risks of Age Verification and Adult Platforms} \label{ssec:privacyav}
Breaches in the adult and intimate-data sector already demonstrated the risks associated with linking identity to sensitive sexual behavior before the current wave of age-verification laws. The 2015 Ashley Madison compromise exposed highly sensitive information on roughly 36 million users, including sexual preferences and relationship status~\cite{ashley-madison-opc-oaic-2016}. Similar incidents affected adult dating and pornography platforms: the 2016 FriendFinder Network breach exposed more than 412 million accounts, the 2012 YouPorn incident leaked chat logs and credentials through a third-party provider, and the CAM4 breach exposed 11 billion records~\cite{greenberg-friendfinder-2016,blake2019age,maris2020tracking,sorn20204exploring}. These cases show that pseudonymous adult platforms and their third-party components can still expose highly sensitive information.

Against this backdrop, children’s rights organizations, civil-society groups, and security and privacy researchers warn that contemporary age-verification tools may reproduce these failures at larger scale. UNICEF notes that document scanning and credit-card checks can create centralized repositories of official identifiers and payment data, giving attackers both identity attributes and evidence of attempted access to sexual or otherwise sensitive content~\cite{unicef_day_2021_ageassurance}. EDRi similarly argues that document-based and biometric schemes are difficult to reconcile with GDPR data-minimization and purpose-limitation principles, and risk infrastructures that link legal identity, browsing habits, and sexual expression~\cite{edri_age_verification_2023}. A 2026 joint statement by security and privacy scientists further argues that age-assurance systems remain difficult to deploy safely at scale, highlighting risks of bypassability, privacy loss, exclusion, and function creep~\cite{ageassurance_scientists_statement_2026}.

Providers often respond with privacy-promoting narratives, such as immediate deletion of images, tokenization of identifiers, or double-anonymity architectures. These assurances are difficult to verify externally and depend on all actors in a complex technical and organizational supply chain behaving correctly. Recent audits and incidents show that collection, retention, and leakage remain significant risks. Investigations of commercial systems deployed on European pornography platforms found external streaming of selfie videos, potentially exposed biometric endpoints, imbalanced facial age-estimation models, governance flaws, cookie-based bypasses, and shared corporate control with porn sites~\cite{aiforensics-agego-2025,aiforensics-ageverif-2025}. Other “privacy-first” providers have been shown to embed third-party tracking and advertising identifiers in apps and web SDKs~\cite{mintsecure-yoti-2025}. The 2024 compromise of an age-verification vendor, the 2025 Discord vendor breach exposing government ID images, and the Tea app leak further illustrate how third-party age and identity checks become attractive extortion targets once deployed at scale~\cite{eff-au10tix-2024,reuters-tea-breaches-2025}.

Taken together, the literature and incident history show that privacy risks in age-verification contexts are not hypothetical: even with tokenization, rapid selfie deletion, or double anonymity, leaks, misuse, and cross-linking of identity and sexual intent must be treated as central threats.

\begin{table}
\centering
\caption{Comparison of related works against our study. \cmarks=\textit{present / explicitly studied}, \xmarks=\textit{absent / not a focus}, \wmarks=\textit{theoretical analysis only}.}
\label{tab:comparison}
\resizebox{\columnwidth}{!}{
\footnotesize
\begin{tabular}{ll ccc}
\toprule
\multicolumn{2}{l}{\textbf{Dimension}} & \textbf{Yao et al.\cite{yao2025easyasch}} & \textbf{Eltaher et al.\cite{eltaher2025digital}} & \textbf{This Work} \\
\midrule
\multirow{4}{*}{\rotatebox{90}{\textit{\textbf{Scope}}}} & 
 Target Domain & Android Apps & Social Media & Sites and Providers \\
 &Jurisdiction & Mixed (18+, 21+) & General (13+, 14+) & EU Mandates (18+) \\
 &Analysis Type & Autom. Detection & Policy Critique & Offensive Pentesting \\
& Scale &  31,750 adult apps &  6 platforms &  154 sites (4 countries)  \\
\midrule
\multirow{3}{*}{\rotatebox{90}{\textit{\textbf{Threats}}}} 
 & Motivated Minor (A1) & \cmarks & \xmarks & \cmarks \\
 & Priv.Consc. Adult (A2) & \xmarks & \xmarks & \cmarks \\
 & AVaaS / Reseller (A3) & \xmarks & \xmarks & \cmarks \\
\midrule
\multirow{6}{*}{\rotatebox{90}{\textit{\textbf{Tech. Assessment}}}} 
 & Fake Media Injection & \wmarks & \xmarks & \cmarks \\
 & Indirect Signals & \wmarks & \cmarks & \cmarks \\
 & Session Replay & \xmarks & \xmarks & \cmarks \\
 & Client-Side Tampering  & \wmarks & \xmarks & \cmarks \\
 & Geofencing Evasion & \cmarks & \xmarks & \cmarks \\
 & Double Anonymity & \xmarks & \xmarks & \wmarks \\
\midrule

\multicolumn{2}{l}{\textit{\textbf{Research Output}}} & Adoption rates & Platform comparison & Adoption rates \\
& & Method Taxonomy & standard compliance & Method Taxonomy \\
& & Detection Tool & Sign-up assessment & Implementation red flags \\
\bottomrule
\end{tabular}
}
\end{table}

\section{Related Works}
The literature distinguishes age verification, age estimation, and broader age assurance, emphasizing that regulatory goals, adversary models, and privacy constraints jointly determine what “works” in practice~\cite{pasquale2020digital,jarvie2024onlineageverification}. A recurring theme is that many deployed controls substitute for higher-assurance digital identity proofing, trading weaker evasion resistance for lower identity binding, while stronger identity checks increase linkability and retention risks unless architectures enforce separation of duties~\cite{nist80063}. Interdisciplinary, policy, and civil-society critiques similarly warn that mandatory age gating can expand surveillance, discourage access in sensitive contexts, and impose privacy and rights costs when unlinkability is not structurally constrained~\cite{apthorpe2025agegating,balsa2023petsmisbehave,stardust2024mandatory,edpb2025statement,edri_age_verification_2023}. 

Empirical work on user behavior further complicates intended protections: parents may help children misstate age, developer compliance processes for child-directed ecosystems are inconsistent, and children’s online behavior is shaped by limited risk awareness, heuristic decision-making, and uneven understanding of privacy and safety risks~\cite{boyd2011whyparents,woodley2025teens,alomar2022developers,zhao2019sillyname}. Policy-oriented work therefore argues that age verification should be situated within broader age-tiered and rights-preserving approaches to child protection, rather than treated as a purely technical access-control problem~\cite{kohlerdauner2025digitalchildprotection}.

Privacy-oriented solutions such as double anonymity align with anonymous credential systems that support selective disclosure of attributes (e.g., “over 18”) with cryptographic unlinkability, as in Idemix, U-Prove, and successors~\cite{camenisch2002idemix,paquin2013uprove,garman2014dac,chase2014kvac}. Token-based approaches such as Privacy Pass similarly support unlinkable issuance and redemption~\cite{davidson2018privacypass}. Recent work extends these ideas to web-native anonymous authentication and non-transferability via device or biometric binding, though binding can reintroduce identifiability and compromise risks~\cite{frolov2025zkcookies,poh2025bbcreds}. In policy, eIDAS~2.0 and the EU Digital Identity Wallet envision selective disclosure via verifiable credentials, but analyses warn that metadata and verifier practices can still reintroduce linkability~\cite{eu2024eidas2,scheffler2025exploringprivacy,chatel2025limitationspets}.

\mypar{Bridging Gaps with Existing Research}
To the best of the authors' knowledge, few works have empirically explored the security of mandated age verification systems. As shown in Table~\ref{tab:comparison}, prior work has examined the adoption of age verification measures in mobile applications intended for adult users ~\cite{yao2025easyasch} and in social media platforms ~\cite{eltaher2025digital}. While these studies operate in domains characterized by heterogeneous age thresholds (e.g., 13+, 18+, or 21+), our work focuses exclusively on pornographic websites in a relatively constrained jurisdiction, namely the European regulatory context, thus addressing a clear legal requirement.
Methodologically, whereas some contributions emphasize automated detection of age verification mechanisms to support regulatory oversight ~\cite{yao2025easyasch}, and others focus on high-level structural critiques ~\cite{eltaher2025digital}, our work focuses on threat modeling and empirical pentesting of age verification architectures, analyzing both the methodologies and their integration within the adult website workflow, to assess whether deployed mechanisms actually achieve their intended protective function.
Our scope further extends beyond content providers to include dedicated solution providers, thereby broadening the attribution of responsibility. We also explicitly incorporate privacy considerations into the threat model, introducing additional actors—such as privacy-aware adults and providers of pre-verified accounts—that are largely absent from prior work. This enables the analysis of misuse scenarios (e.g., account sharing or resale) that are typically overlooked.
Finally, our conclusions diverge from those of prior work. We empirically assess that most age verification methodologies, including biometric age estimation and document-plus-selfie workflows, are largely ineffective in practice, and often impractical due to jurisdictional heterogeneity, despite being frequently presented as robust in the literature ~\cite{yao2025easyasch, eltaher2025digital}.

\section{Motivation}

Age-verification mechanisms deployed on pornographic websites create a security and privacy tradeoff that remains poorly understood in practice. To reduce minors' access to adult content, these mechanisms often require users to disclose sensitive identity-related attributes in a context where linkage to browsing behavior is especially privacy-sensitive. 
However, regulators need to prioritize auditable compliance and child-protection objectives, while websites and third-party providers also seek to minimize friction, operational costs, user abandonment, and liability. As a result, deployed systems may favor mechanisms that are easy to integrate and defensible from a compliance perspective, without necessarily offering robustness against motivated underage users, privacy-conscious adults, or actors interested in monetizing reusable proofs of age.

The question that shapes our research is therefore whether current deployments provide meaningful protection against realistic circumvention attempts or satisfy compliance requirements while introducing new privacy and abuse risks.

Although \textit{privacy} risks associated with age-verification systems have been repeatedly documented (See Section~\ref{ssec:privacyav}), to the best of our knowledge, no prior work has conducted a systematic, large-scale \textit{security} assessment of regulation-mandated age-verification mechanisms and their real-world implementations on adult websites. Nonetheless, prior studies suggest that many deployed age-verification approaches, such as biometric age estimation, are susceptible to practical circumvention and adversarial workarounds~\cite{yao2025easyasch}. Motivated by these observations, we examine how age-verification mechanisms are currently deployed by adult websites and third-party providers, and whether their deployed security properties align with realistic adversary models. We structure our investigation around the following research questions:

\mypar{RQ1} \textit{Despite regulatory requirements, what fraction of pornographic websites currently deploy age verification?}

\mypar{RQ2} \textit{Which techniques are predominantly adopted by compliant websites and third-party providers across jurisdictions?}

\mypar{RQ3} \textit{To what extent do compliant websites and providers implement robust mechanisms and deployment strategies?}

\mypar{RQ4} \textit{Do deployments reflect a realistic threat model?}
\smallskip

We conclude by interpreting our findings as an empirical assessment of current deployment practices, identifying which classes of mechanisms appear more promising, and discussing whether and under what conditions their desirable security and privacy properties can be achieved in practice. By grounding a politically and socially contested policy debate in empirical evidence, our goal is not to argue for or against age assurance for adult content per se. Rather, we aim to answer whether current approaches and deployments provide protective benefits commensurate with the privacy risks they introduce, or they primarily operate as an X-rated compliance theater.
%%%%%%%%%%%%%%%%%%%%%%%%%%%%%%%%%%%%%%%%%%%%%%%%%%%%%%%%%%%%%%%%%%%%%%%%%%%%%%%%
\section{Threat Model}
\label{sec:threatmodel}
We categorize adversaries in the age-verification ecosystem for adult websites into two classes: (A) those who bypass or circumvent verification and (B) those who access or exploit sensitive user data, further distinguishing them by objectives and technical capabilities.

\mypar{A1 - Motivated Underage User} Such an adversary is the main target of European regulations. Multiple works in the literature~\cite{pasquale2020digital} (and common knowledge) support the existence of this attacker, who can be identified as a minor seeking to bypass age verification barriers to access adult content. The motivation of such an attacker will likely push them towards simple workarounds, such as falsifying self-declaration forms, borrowing adult credentials (e.g., parents' credentials), using VPNs~\cite{unicef2025savvyteens,eltaher2025digital}, or other low-cost approaches. 

\mypar{A2 - Privacy-conscious Adult User} Multiple studies show a lack of trust towards data handling of age verification mechanisms~\cite{jarvie2024onlineageverification, Ofcom2022}. This adversary is a legally adult user, entitled to consume the content, but aims to preserve anonymity and avoid leaving a trail of their content consumption due to concerns about privacy and data misuse. These users, in addition to potentially using unsanctioned and less trustworthy sources, will act similarly to the motivated underage user, using workarounds such as VPNs. However, they have access to more resources, may be more inclined to spend money to achieve their goal, and cannot rely on borrowed credentials. 

\mypar{A3 - Fraudulent Identity Provider} Impersonation-as-a-Service (ImpaaS) markets are well documented in prior work~\cite{campobasso2020impersonation}, and deepfake-enabled fraud has already been observed in other domains~\cite{trifonova2024misinformation}. We model this adversary as an organized black-market actor that offers bypass solutions for age and identity checks, i.e., an AVaaS (Age-Verification-as-a-Service) provider. This adversary capitalizes on the demand created by mandatory verification: once age authentication is required, a profitable niche emerges for supplying geofencing bypasses, pre-verified accounts, forged or replayable identity artifacts, and, in some cases, access to real individuals' credentials to users who are unwilling or unable to complete verification legitimately. We also include in this category ideologically motivated actors (e.g., hacktivists) who may develop and distribute comparable bypass tooling to resist perceived population-level control, rather than for direct financial gain. This attacker's capabilities are significant in relation to the previous two: they may employ automation and AI tools to mass-create verified accounts via deepfakes or alter ID photos, as well as stolen or acquired personal data. They can be considered technically competent in approaching the exploitation of age verification mechanisms. Their goal differs from the previous two, as they need reproducible and shareable solutions that can be offered to their clients.

\mypar{B1 - External Privacy Threat} This attacker is a classical external adversary, a criminal group that targets the data repositories and infrastructure of age-verification systems or adult websites to steal sensitive information. In our context, the prize for this attacker is the trove of identity data (scans of IDs, selfies, biometric hashes, etc.) collected to enforce age restrictions on adult sites.
The literature strongly justifies this threat’s inclusion as security experts and privacy scholars have cautioned that centralized identification data for porn access would become remunerative targets for attackers~\cite{goldman2025segregate}.

\mypar{B2 - Insider Privacy Threat} This adversary represents insider threats within the ecosystem, including providers, adult websites, or integrated third parties, which may compromise the trust placed in them and breach user privacy. This final adversary is already acknowledged by some European states' regulators, who call for a "double anonymity" architecture that would not allow the adult website or the age verification provider to access both identity and intent data. In threat modeling terms, the porn site and the verification service could collude (or be a single entity) to deanonymize users for profit or surveillance. The capabilities here are those any large platform has: extensive analytics infrastructure, the ability to correlate accounts, and the incentive to monetize data.

\subsection{Required Security and Privacy Properties}
We identify three properties that any age verification architecture for adult content access should satisfy, obtained by assessing adversaries' interests and capabilities:

\mypar{P1 - Robustness Against Evasion}
The system must guarantee a reliable determination of age that resists bypass strategies. This includes preventing false self-attestation, blocking the use of third-party credentials, and eliminating reliance on client-side checks that can be altered or revoked by users. 

\mypar{P2 - Non-Shareability and Anti-Replay Guarantees}
Verification outputs must be intrinsically bound to the individual undergoing authentication, ensuring that they cannot be transferred or sold. This entails replay protection, prevention of token cloning, and defenses against the automated creation of large-scale synthetic identities. The system should ensure that a legitimate verification event cannot be reproduced, thereby preventing unauthorized parties from gaining access.

\mypar{P3 - Unlinkability of Identity and Intent}
No single party should learn both the user’s sensitive identity data and the specific site or content category they access. Privacy requires separating age/identity validation from content access.

\section{Age Verification Deployment}
\label{sec:securityeval}

To assess the current state of age-verification deployment in Europe, this section presents an empirical analysis of age-verification systems for online adult content. We first describe how we compiled and accessed a cross-country dataset of 154 high-traffic adult websites in Italy, Germany, France, and the United Kingdom. We then measure age-verification adoption and characterize the verification methods and third-party providers (RQ1--RQ2).

\subsection{Data Collection Methodology}
\label{ssec:ecosystem}

To construct our dataset of adult websites, we conduct a country-level market scan\footnote{via https://semrush.com, filtering by adult industry and location, resulting in the most viewed websites in November 2025.} to identify the most visited platforms in the adult category across Italy, Germany, France, and the United Kingdom. For each country, we extract the top 100 websites by traffic, then consolidate the four national lists into a single dataset, removing duplicates. This process yields an initial pool of 230 unique adult websites. Following a screening phase to exclude platforms not aligned with the scope of our analysis (we include websites that provide explicit pornographic material, excluding marketplaces, dating services, or aggregators), we finalize a curated sample of 154 websites that form the basis of our study.

We evaluate each website from the three EU countries that have already introduced age-verification obligations, but under different legal frameworks. To provide a comparison with a more mature regulatory context, we also include the United Kingdom, which has incorporated age-verification requirements earlier and under separate, yet comparable, legislation. For each country, we route traffic through a local VPN exit node to ensure the website is accessed under that jurisdiction's conditions.\footnote{To assess whether websites treated commercial VPN exit nodes differently from non-VPN connections, we accessed them both through an Italian VPN exit node and from a clean Italian IP address. We observed no differences in the resulting age-verification behavior.}

We then document for each site: (a) the presence of age-verification measures, (b) any integrated third-party age-assurance providers, (c) the verification methods they offer, and (d) whether the website or provider enables linking the verification to a user account. See the Open Science section in appendix for details regarding our dataset.

We manually inspected each site using a shared annotation spreadsheet and codebook covering AV presence, provider/workflow, methods exposed by country, account binding, reusable proofs, and double-anonymity claims. Two researchers piloted the codebook, performed the annotations, and unresolved cases were adjudicated by a third researcher.

\begin{table*}
\caption{Deployment of each provider in websites upon access from the given country. Please note that multiple providers may be available on a single website.}
\centering
\resizebox{\textwidth}{!}{
\begin{tabular}{l c c c c c c c c c c c c c c c c c c c }
\toprule
\textbf{Nation} & \textbf{AV Present} & \textbf{Double Anonymity} &
\textbf{\yoti} &
\textbf{\ageverif} &
\textbf{\agego} &
\textbf{\verifymy} &
\textbf{\friendlyid} &
\textbf{\oneid} &
\textbf{\gotaca} &
\textbf{\paymentico} &
\textbf{\anonymage} &
\textbf{\veratad} &
\textbf{\pleenk} &
\textbf{\ondato} &
\textbf{\gocam} &
\textbf{\agerify} &
\textbf{\verifyage} &
\textbf{Internal} \\
\midrule
UK      & 69/154 & 40/69 & 31 & 37 & 11 & 9 & 2 & 5 & 18 & 2 & 0 & 1 & 0 & 1 & 5 & 3 & 1 & 6\\
France  & 38/154 & 29/38 & 30 & 28 & 3 & 1 & 1 & 0 & 16 & 0 & 1 & 1 & 1 & 1 & 1 & 0 & 0 & 0\\
Germany & 7/154  & 3/7   & 7  & 0  & 0 & 1 & 1 & 0 & 0  & 0 & 0 & 1 & 0 & 0 & 0 & 0 & 0 & 0\\
Italy   & 4/154  & 4/4   & 4  & 0  & 0 & 0 & 0 & 0 & 0  & 0 & 0 & 1 & 0 & 0 & 0 & 0 & 0 & 0\\
\bottomrule
\end{tabular}}
\label{tab:av-state-providers}
\end{table*}

\subsection{Age Verification Presence}

\begin{table}
\caption{Number of websites offering each verification method in the assessed countries. Counts are non-exclusive: a website may offer multiple methods.}
\centering
\resizebox{\columnwidth}{!}{
\begin{tabular}{l c c c c c c c}
\toprule
\textbf{Nation} &
\textbf{Age Estimation} &
\textbf{ID} &
\textbf{Email} &
\textbf{SMS} &
\textbf{Credit Card} &
\textbf{Physical Ticket} &
\textbf{eID}\\
\midrule
UK      & 60 & 14 & 17 & 5 & 28 & 0 & 0\\
France  & 36 & 5  & 8  & 0 & 0  & 4 & 0\\
Germany & 6  & 5  & 0  & 0 & 0  & 0 & 0\\
Italy   & 3  & 2  & 0  & 0 & 0  & 0 & 0\\
\bottomrule
\end{tabular}}
\label{tab:av-state-mechanisms}
\end{table}

\begin{table}
\caption{Available verification methods per provider. Websites using a provider may not support all the provider's methods.}
\centering
\resizebox{\columnwidth}{!}{
\begin{tabular}{l c c c c c c c}
\toprule
\textbf{Provider} &
\textbf{Age Estimation} &
\textbf{ID} &
\textbf{Email} &
\textbf{SMS} &
\textbf{Credit Card} &
\textbf{Account} &
\textbf{Physical Ticket}\\
\midrule
\ageverif    & \checkmark &   & \checkmark &   & \checkmark & \checkmark & \checkmark \\
\agego       & \checkmark &   &   &   & \checkmark & \checkmark &   \\
\gotaca      & \checkmark &   &   &   &   & \checkmark &  \\
\yoti        & \checkmark & \checkmark &   &   &   & \checkmark &   \\
\verifymy    & \checkmark & \checkmark & \checkmark &   &   & \checkmark &   \\
\friendlyid  & \checkmark &   &   &   &   &   &   \\
\oneid       &   &   &   & \checkmark & \checkmark &   &   \\
\paymentico  &   &   &   &   & \checkmark &   &   \\
\anonymage   &   & \checkmark &   &   &   & \checkmark &   \\
\veratad     &   & \checkmark &   &   &   & \checkmark &   \\
\pleenk      & \checkmark &   &   &   &   & \checkmark &   \\
\gocam       & \checkmark & \checkmark &   &   &   &   &   \\
\ondato      &   & \checkmark &   &   &   &   &   \\
\agerify     & \checkmark & \checkmark &   &   &   &   &   \\
\verifyage   & \checkmark & \checkmark &   &   &   &   &   \\
Internal system* & \checkmark &   &   &   & \checkmark &   &   \\
\bottomrule
\multicolumn{8}{l}{*Internal system refers to all websites implementing verification without relying on a third-party provider.}
\end{tabular}}
\label{tab:av-provider-mechanisms}
\end{table}

\begin{figure}[h]
\centering
\begin{tikzpicture}
\begin{axis}[
    width=\linewidth, height=5cm,
    xlabel={Visit decile (1 = lowest traffic, 10 = highest traffic)},
    ylabel={Share of sites with $\ge$1 measure},
    ymin=0, ymax=0.7,
    ymajorgrids=true,
    legend pos=north west,
]
\addplot table[x=decile, y=UK_has_av, col sep=comma]{images/av_by_visit_decile.csv};
\addlegendentry{UK}

\addplot table[x=decile, y=France_has_av, col sep=comma]{images/av_by_visit_decile.csv};
\addlegendentry{France}

\addplot table[x=decile, y=Germany_has_av, col sep=comma]{images/av_by_visit_decile.csv};
\addlegendentry{Germany}

\addplot table[x=decile, y=Italy_has_av, col sep=comma]{images/av_by_visit_decile.csv};
\addlegendentry{Italy}

\end{axis}
\end{tikzpicture}
\caption{Prevalence of age verification (AV) by traffic decile.}
\label{fig:deciledistribution}
\end{figure}
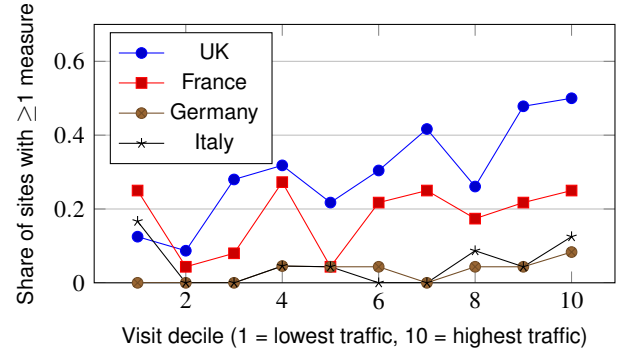

Tables~\ref{tab:av-state-providers},~\ref{tab:av-state-mechanisms}, and~\ref{tab:av-provider-mechanisms} summarize observed age-verification providers, method prevalence, and supported provider mechanisms, while Figure~\ref{fig:deciledistribution} relates deployment to traffic share. Deployment is highly uneven: the UK is most mature (69/154 sites), followed by France (38/154), Germany (7/154), and Italy (4/154).

\mypara{Providers} We identify 15 verification providers used by the analyzed websites, namely AgeGo, Agerify, AgeVerif, Anonymage, FriendlyID, Gataca, Go.cam, OneID, Ondato, Paymentico, Pleenk, Veratad, VerifyAge, Verifymy, and Yoti. 
The registered corporate geography of providers is distributed across Cyprus, the Czech Republic, France, Lithuania, Portugal, Spain, the UK, and the US. This geography should not be interpreted as the country in which a provider is deployed: most providers operate across multiple jurisdictions and are integrated by websites serving users in countries different from the provider's apparent corporate base.
As our objective is not to perform a comparative evaluation of these providers, we anonymize them in the remainder of the paper and refer to each as \textit{VPX} (Verification Provider~X).

\begin{answer}
\textbf{Answer to RQ1:} We find that the majority of websites do not implement age-verification mechanisms in EU countries, despite such measures being legally mandatory.
\end{answer}

Traffic share only partially explains adoption. In the UK, age verification spans most deciles but remains non-universal even among the most visited sites, consistent with stronger scrutiny of large operators. France concentrates slightly more deployment among higher-traffic sites, whereas Germany and Italy remain near the floor across deciles. Figure~\ref{fig:deciledistribution} further shows that adoption is only partially concentrated among more-visited sites, meaning that exposure to age verification varies substantially across the popularity distribution.

Provider adoption is similarly concentrated. We find that \ageverif and \yoti dominate in the UK and France, while Germany and Italy are almost exclusively anchored to \yoti, with only isolated alternatives. The recurrence of the same providers across jurisdictions is consistent with platform-level integration choices playing an important role. This is relevant because we observe cases where deployed workflows appear potentially misaligned with certain national constraints, such as account creation being present in contexts where it should be restricted.

Mechanism choice largely converges on selfie-based age estimation, with ID document verification a distant second. As Table~\ref{tab:av-state-mechanisms} shows, age estimation is the most common method in all four countries, while ID-based verification appears less frequently. Indirect signals are jurisdiction-dependent: credit-card checks are common in the UK but absent elsewhere, email-based checks appear in the UK and France, SMS is rare, and physical tickets are only observed in France. Although providers often support multiple modalities (see Table~\ref{tab:av-provider-mechanisms}), websites typically expose a narrower, country-dependent subset, consistent with regulatory and operational constraints. We observe no use of national eID or eIDAS-style credentials, despite their prominence in policy discussions. This absence is likely explained by the perceived mismatch between the strong identity binding inherent in such systems and the privacy risks associated with accessing adult content.

Finally, across all nations, a varying fraction of websites offer at least one double-anonymity option, under which the age-verification provider is purportedly unable to infer the content accessed by the user, while the content provider is unable to access the user’s real-world identity. In practice, however, these guarantees are largely non-verifiable from the user’s perspective, as their enforcement depends on opaque implementations and data-handling practices.

\begin{answer}
\textbf{Answer to RQ2:} Age estimation emerges as the main verification technique. ID-based verification is the second most common approach, while softer signals, such as email, SMS, or credit card–based checks, are adopted more sporadically. Notably, none of the surveyed providers offer verification mechanisms based on digital identity frameworks (e.g., national eIDs or eIDAS-compliant credentials). 
\end{answer}

\section{Exploiting Verification Weaknesses}
We examine practical strategies to bypass age verification by completing verification without valid credentials. We evaluate all methods except physical tickets, as we found no vendors offering them, despite checking provider-listed French shops.
Security testing was primarily manual because workflows varied across sites and to avoid large-scale automated interaction with live services. We used limited automation only to confirm that selected bypasses could be scripted, but scaling is outside the scope of this work. Before adversarial testing, researchers completed the relevant flows legitimately to confirm baseline functionality.

\subsection{Age Estimation}  
% =====================================================================
% (a) AGE ESTIMATION  -- spoofed camera input
% =====================================================================
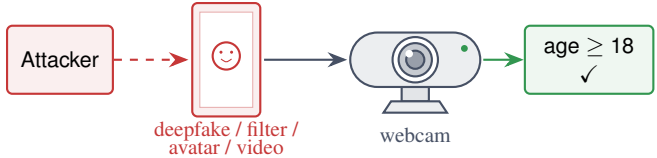
\begin{figure}[h]
\centering

\begin{tikzpicture}
  % attacker
  \node[attbox, minimum width=1.4cm, minimum height=0.9cm] (att) at (0,0)
       {\shortstack{Attacker}};
 
  % phone displaying a fake face
  \phoneframe{(2.2,0)}{0.45}{0.75}
  \tinyface{(2.2,0.05)}{att}
  \node[font=\footnotesize, att] at (2.2,-0.95) {deepfake / filter /};
  \node[font=\footnotesize, att] at (2.2,-1.15) {avatar / video};
 
  % real webcam icon
  \camera{(4.7,0)}{1}
  \node[font=\footnotesize, sys] at (4.7,-1) {webcam};
 
  % accepted
  \node[okbox, minimum width=1.7cm, minimum height=0.9cm] (ok) at (7,0)
       {\shortstack{age $\geq$ 18\\ \checkmark}};
 
  \draw[attflow] (att) -- (1.7,0);
  \draw[flow]    (2.7,0) -- (3.8,0);
  \draw[okflow]  (5.6,0) -- (ok);
\end{tikzpicture}

\caption{Overview of age estimation attacks methodology.}
\label{fig:ageestimation}
\end{figure}

Selfie-based age estimation is commonly augmented with anti-fraud controls intended to detect presentation and injection attacks. Beyond basic face detection and age-regression models, providers typically deploy (i) liveness checks, either passive (e.g., texture cues, micro-movements, eye-blink dynamics) or active (e.g., prompted head turns, pose changes, or expression challenges); (ii) screen/presentation-attack detection, including heuristics based on Moir\'e patterns, re-capture artifacts, or focus/lighting inconsistencies~\cite{screendetect}; (iii) challenge-response illumination (face flashing), where randomized on-screen color sequences are matched against measured facial reflections~\cite{tang2018faceflashingsecureliveness}; (iv) audio-visual challenges (e.g., spoken prompts with lip-sync checks)~\cite{Mandalapu_2021}; and (v) platform integrity signals, such as emulator/root detection or device attestation, to constrain the environment from which age verification is performed. 

\mypar{Methodology}
To characterize the robustness of deployed age verification pipelines, we evaluate whether an adversary can complete age verification while controlling the camera input, see Fig.~\ref{fig:ageestimation}. We inject synthetic visual input using three attack channels and eight media types representative of low- to moderate-effort abuse.
We consider three attack channels that differ in where the adversary controls the camera stream.
We select these three attack channels to reflect a realistic escalation path for controlling the camera input, following prior work and public analyses of presentation and video-injection attacks against remote identity and liveness checks~\cite{carta2023video,carta2022video}.

\noindent (1) \textit{Presentation attacks}: the adversary displays visual content on a secondary screen (e.g., a laptop) and points the verification device's camera at that screen. For this experiment, we perform age verification on a mobile device while displaying the media on a second device and capturing it via camera.

\noindent (2) \textit{Emulator-based attacks}: the adversary runs the verification workflow in a mobile emulation environment and supplies a synthetic camera stream in place of a physical camera feed. It models attackers who can satisfy mobile-only requirements while retaining full control over video input. Here, we perform age verification within a mobile emulation environment on a laptop, using the candidate media as the camera input.

\noindent (3) \textit{Rooted-device attacks}: the adversary compromises a smartphone and replaces camera frames at the OS/API boundary with pre-recorded or synthetic frames. It models attackers who retain the outward appearance of a real device (sensors, OS signals) while controlling visual input. We perform age verification on a rooted smartphone while supplying the candidate media as the camera input. 

We evaluate eight media types to assess the robustness of the validation process. \textit{Images (I)} are single high-resolution portraits of adults. \textit{3D models (3D)} are interactive face meshes rendered in a browser, manipulated (rotation/zoom) to approximate head motion and distance changes. \textit{Social-media-style filters (S)} are real-time face effects (e.g., ageing overlays, beard overlays, and full-face masks) applied to a live webcam feed. \textit{Deepfakes (DF)} are real-time face-swap videos produced using publicly available software~\cite{DFX}. \textit{Videogame avatars (VG)} are realistic 3D characters created in commercial games (inZOI and Cyberpunk2077), captured using in-game photo/camera modes and natural animation. Finally, \textit{Videos (V)} are short, pre-recorded clips of adults maintaining eye contact and performing natural head and distance movements consistent with typical liveness prompts (while we used our own, such videos are now commonly available online on video streaming platforms).
We first complete each provider’s verification procedure to document the required interaction, such as prompts for pose changes or distance adjustments. We then test each attack channel and media type. As live age estimation systems may reject inputs for transient reasons, the results in Table~\ref{tab:av-media-vs-provider} report whether a bypass was obtained for a provider/media/channel combination within the first three attempts.

\mypar{Results}
The aggregated outcomes are shown in Table~\ref{tab:av-media-vs-provider}. 10 out of 15 providers support selfie-based age estimation. Across these providers, we observe diverse interactive prompts: 1 out of 10 (\ageverif) requires explicit head rotation during the procedure, 3 out of 10 (\gocam, \agerify and \verifyage) require a change in facial expression, while the remaining 6 out of 10 rely primarily on distance adjustments. In particular, 2 out of these 6 (\agego and \friendlyid) request the user to move away from the camera, and all 6 ultimately require the user to move closer and remain still to finalize capture.

\noindent\textit{Attack-channel effectiveness.}
Presentation attacks are the least effective attack channel (success rate 5/10), consistent with widespread deployment of screen/presentation-attack detection. Notably, \ageverif , \agerify and \verifyage do not appear to detect presentation attacks across any tested media. \gotaca and \gocam exhibit inconsistent behavior: they reject some media but are bypassed by others, such as 3D models and pre-recorded videos.

Emulator-based attacks achieve a success rate of 9/10. \yoti effectively prevents completion under emulation: it allows the procedure to start but fails to accept the injected face stream. The remaining providers do not consistently enforce device-environment constraints, resulting in outcomes that are largely dependent on the media type.

Rooted-device attacks achieve a success rate of 7/10: \gotaca constrains the workflow to a specific browser environment, which reduces the effectiveness of our camera replacement; \friendlyid and \pleenk deploy a randomized color-flash challenge~\cite{tang2018faceflashingsecureliveness}, which prevents reuse of prerecorded videos in our setting. However, both remain bypassable in our emulator-based setting, indicating that the flash-based defense, while promising, is not sufficient when the attacker can control the live stream end-to-end.

%\smallskip
\noindent\textit{Media-type effectiveness.}
Videogame avatars and Images yield the lowest bypass rate (4/10). Despite high visual realism, avatars remain distinguishable from real captures, and frame-rate or animation artifacts likely reduce their plausibility under behavioral analysis. Conversely, static media fails when pipelines leverage micro movement or blink cues.
Interestingly, filters, trivial to obtain by all attackers, produce the highest bypass rate (9/10); the only provider that resists filter-based bypasses is \yoti.
Videos and \textit{deepfakes} achieve similar aggregate success rates (8/10), but against different subsets of providers, suggesting heterogeneous acceptance criteria and sensitivity to artifacts such as boundary inconsistencies and motion smoothness.
3D models achieve an intermediate success rate (7/10), plausibly limited by the absence of subtle facial dynamics, lack of torso, and unnatural backgrounds.

\noindent\textit{Overall robustness.}
No provider is robust to all tested attack channels and media types. Conversely, no single attack channel is universally effective: successful bypass requires selecting the channel and media type that best match the provider's deployed defenses. We note that the evaluated bypasses are automatable (e.g., scripted session orchestration and media replay), which lowers the marginal cost of repeated attempts once an attacker identifies a strategy.

Note that, although \ageverif requires head rotation as an active liveness prompt, this control is insufficient to prevent bypass, suggesting that it does not robustly validate the visual consistency of the face under motion.
Furthermore, although \gocam, \agerify, and \verifyage require a change in facial expression (a measure intended to prevent the use of pre-recorded videos) this mechanism proves ineffective. By consistently relying on the same action prompts, these systems are easily predictable.
The internal system lacks a liveness mechanism and is vulnerable to all attack vectors, regardless of the media type used. 
%Its adoption appears to serve as a regulatory compliance measure, rather than constituting a genuinely effective mechanism to prevent minors from accessing harmful content.
Finally, \ageverif , \gocam , \agerify and \verifyage do not exhibit evidence of liveness checks that leverage movements or blinking, consistent with their susceptibility to low-dynamics media.

  \begin{table}
  \centering
  \caption{Successful bypass channels per provider and media type.
  Codes: P = presentation attack (screen replay), E = emulator-based camera injection, R = rooted-device camera hook. An empty cell indicates that no bypass was obtained for that provider media/attack combination.}
  \footnotesize
  \resizebox{\columnwidth}{!}{
  \begin{tabular}{lcccccc}
    \toprule
    Provider
      & Image & 3D model & Snapchat filter & Deepfake & Videogame avatar & Video \\
    \midrule
    \ageverif   & P,E,R & P,E,R & P,E,R & P,E,R & P,E,R & P,E,R \\
    \agego      &       & E     & E,R   &       &       & E,R   \\
    \gotaca     &       & P,E   & E     & E     &       & P,E   \\
    \yoti       &       &       &       & R     &       & R     \\
    \verifymy   &       & E     & E     & E     &       & E,R   \\
    \friendlyid &       &       & E     & E     &       &       \\
    \pleenk     &       &       & E     &       &       &       \\
    \gocam      & P,R   & P,R   & P,E,R & E,R   & P,R   & P,E,R \\
    \agerify    & P,E,R & P,E,R & P,E,R & P,E,R & P,E,R & P,E,R \\
    \verifyage  & P,E,R & P,E,R & P,E,R & P,E,R & P,E,R & P,E,R \\
    \bottomrule
  \end{tabular}}
  \label{tab:av-media-vs-provider}
\end{table}

\mypar{Adversary relevance}
Presentation attacks and filter-based media are realistic for motivated underage users (A1), as they require only commodity devices and widely available media-editing tools. Emulator-based and rooted-device attacks require more technical effort and are therefore more plausible for privacy-conscious adults (A2) or technically capable users. Once a working provider-specific configuration is identified, however, the same attack logic can be packaged or automated, making these channels relevant to AVaaS actors (A3).

\subsection{Document Upload} 

% =====================================================================
% (b) DOCUMENT UPLOAD  -- forged / recaptured ID
% =====================================================================
\begin{figure}[h]
    \centering
\begin{tikzpicture}
 
  \node[attbox, minimum width=1.4cm, minimum height=0.9cm] (att) at (0,0)
       {Attacker};
 
  % fake ID card
  \begin{scope}[shift={(2.3,0)}]
    \draw[attbox] (-0.7,-0.45) rectangle (0.65,0.4);
    \draw[fill=att!20, draw=att] (-0.6,-0.3) rectangle (-0.2,0.3);
    \tinyface{(-0.4,0)}{att}
    \draw[att, thick] (-0.05,0.25) -- (0.6,0.25);
    \draw[att, thick] (-0.05,0.05) -- (0.6,0.05);
    \draw[att, thick] (-0.05,-0.15) -- (0.4,-0.15);
    \node[att, font=\tiny\bfseries, rotate=-15] at (0.25,-0.35) {FAKE};
  \end{scope}
  \node[font=\footnotesize, att] at (2.3,-0.85) {forged / online / AI};
 
  % website (browser) -- replaces "upload + verifier"
  \browser{(4.7,0)}{0.8}{0.65}
  \node[font=\footnotesize, sys] at (4.7,-0.9) {website};
 
  % accepted
  \node[okbox, minimum width=1.7cm, minimum height=0.9cm] (ok) at (7,0)
       {\shortstack{age $\geq$ 18\\ \checkmark}};
 
  \draw[attflow] (att) -- (1.55,0);
  \draw[flow]    (3.05,0) -- (3.8,0);
  \draw[okflow]  (5.6,0) -- (ok);
\end{tikzpicture}
    \caption{Overview of document upload attacks methodology.}
    \label{fig:documentupload}
\end{figure}
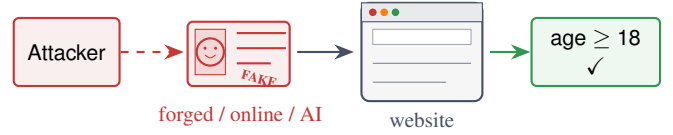

This experiment's objective is to test whether the verification mechanisms are robust against forgery of identity documents, with the goal of assessing whether the provider performs anti-fraud checks (as opposed to merely extracting a date of birth).

\mypar{Methodology}
We first identify the providers that offer age verification via identity-document upload on at least one tested adult website in at least one of the four countries in scope (Italy, Germany, France, and the United Kingdom). For each eligible provider, we submit clearly fake identity documents, not associated with any real-world identity, to assess their robustness.
When the workflow requires only the front of the document, we use a clearly fake driving license, obtained from publicly available online material. When the workflow requires both front and back images, we use an AI-generated driving license to ensure that a matching pair of images is available for upload. When the provider requires a photograph of the document (rather than a file upload), we use a second device to photograph the document displayed on a laptop screen, allowing us to also probe for the presence of basic presentation-attack checks (e.g., screen re-capture detection).
We do not craft high-fidelity forgeries nor use legitimate identity data in the crafted documents (e.g., valid ID codes); rather, we intentionally create recognizable fakes to test whether the pipeline rejects implausible inputs.

\mypar{Results}
Table \ref{tab:av-id-features} summarizes the document-upload features across providers.
Overall, 8 out of 15 operators support age verification via document upload. Among these, 3 out of 8 (\veratad , \ondato and \anonymage) require both the front and back of the document; in our experiments, this appears to trigger a basic cross-image consistency check. Requiring both sides raises the bar against opportunistic misuse of single images found online (as it is less common to find both sides of the same document), but it does not eliminate fraud risks in the presence of stolen-document markets. We also observe that 5 out of 8 operators (except \veratad, \ondato and \verifymy) allow users to either upload an existing image from local storage or capture a photo at verification time. Allowing direct file upload removes an additional friction step and can simplify abuse by adversaries who can source document images without physical access to the document. Two operators (\verifymy and \veratad) additionally require a selfie-based face match between the user and the document portrait. This step can mitigate straightforward sharing of third-party documents (e.g., minors using an adult’s ID), but it materially increases data sensitivity and, in some jurisdictions (e.g., Italy), may not be permissible in this context.
Finally, 4 out of 8 operators (\yoti , \gocam , \agerify and \verifyage) are ultimately vulnerable to fake document injection. \yoti does not reject the fake ID image. The others do not reject the ID and accept a screen capture, suggesting the absence of anti-fraud checks for presentation attacks in this workflow.

Beyond the document-upload pipeline itself, we observe heterogeneity in platform hardening. Although \verifymy and \veratad are vulnerable to emulator-based attacks in other pipelines, they detect the camera-injection attempt required to supply an altered video stream for face matching.

\mypar{Adversary relevance}
Low-effort document misuse, such as reusing an adult's document or submitting an online-available fake, is within reach of motivated underage users (A1). Screen recapture or pairing with face-matching workflows is more realistic for privacy-conscious adults (A2) and especially AVaaS actors (A3), who may produce digital versions of fake documents to distribute.

\begin{table}
\caption{Document upload verification and attack results.}
\centering
\footnotesize
\resizebox{\columnwidth}{!}{
\begin{tabular}{l c c c c c c c }
\toprule
\textbf{Provider} & \multicolumn{3}{c}{\textbf{Requirements}} & \textbf{Vulnerable to fake ID} \\

 &
\textbf{Front/Back} &
\textbf{Selfie required} &
\textbf{Photo upload} &
 \\
\midrule
\yoti        &  &  &  \checkmark & \checkmark \\
%\emblem      &  &  &  &  \checkmark\\
\verifymy    &  &  \checkmark &  &   \\
\anonymage   &  \checkmark &  &  \checkmark &   \\
\veratad     &  \checkmark &  \checkmark &  &   \\
\ondato      &  \checkmark &  &  &  & \\
\gocam       &  &  &  \checkmark & \checkmark \\
\agerify     &  &  &  \checkmark & \checkmark \\
\verifyage   &  &  &  \checkmark & \checkmark \\
\bottomrule
\end{tabular}}
\label{tab:av-id-features}
\end{table}

\subsection{Email, SMS, and Credit Cards} 

% =====================================================================
% (c) EMAIL / SMS / CREDIT CARD  -- disposable credentials
% =====================================================================
\begin{figure}[h]
    \centering
\begin{tikzpicture}
  \node[attbox, minimum width=1.4cm, minimum height=0.9cm] (att) at (0,0)
       {Attacker};
 
  % three disposable credentials stacked
  \begin{scope}[shift={(2.2,0)}]
    \draw[attbox] (-0.55,0.55) rectangle (0.55,1.15);
    \draw[att]   (-0.55,1.15) -- (0,0.75) -- (0.55,1.15);
    \node[font=\footnotesize, att] at (0,0.4) {temp email};
 
    \draw[attbox] (-0.3,-0.25) rectangle (0.3,0.25);
    \draw[fill=white, draw=att!40] (-0.2,-0.17) rectangle (0.2,0.17);
    \node[font=\footnotesize, att] at (0,-0.45) {virtual SIM};
 
    \draw[attbox] (-0.55,-1.15) rectangle (0.55,-0.7);
    \draw[fill=att!25, draw=att] (-0.5,-0.85) rectangle (-0.2,-0.75);
    \draw[att, thick] (-0.15,-0.82) -- (0.5,-0.82);
    \node[font=\footnotesize, att] at (0,-1.3) {shared card};
  \end{scope}
 
  % website (browser) -- replaces "indirect-signal check"
  \browser{(4.7,0)}{0.8}{0.7}
  \node[font=\footnotesize, sys] at (4.8,-0.9) {website};
 
  % accepted
  \node[okbox, minimum width=1.7cm, minimum height=0.9cm] (ok) at (7,0)
       {\shortstack{age $\geq$ 18\\ \checkmark}};
 
  \draw[attflow] (att) -- (1.75,0);
  \draw[flow]    (3.05,0.85)  .. controls (3.4,0.4)  .. (3.8,0.15);
  \draw[flow]    (2.75,0)     -- (3.8,0);
  \draw[flow]    (3.05,-0.9)  .. controls (3.4,-0.4) .. (3.8,-0.15);
  \draw[okflow]  (5.6,0) -- (ok);
\end{tikzpicture}
\label{fig:indirect}
\caption{Overview of indirect channels attacks methodology.}
\end{figure}
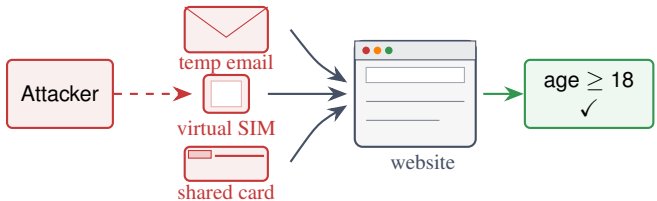

While obtaining valid email addresses, phone numbers, or credit cards from third parties is feasible, we instead examine whether temporary credentials suffice, enabling automated account creation and large-scale redistribution.

\mypar{Methodology}
For \textit{email-based} verification, we first identify the operators that offer this method on at least one of the tested websites. For each such operator, we attempt to complete the flow using temporary email addresses generated through common disposable email providers, namely temp-mail, guerrillamail, internxt, and mail.tm. When the attempt fails, we repeat the procedure using a personal email address.

For \textit{SMS-based} verification, we similarly identify the operators that expose this option on at least one website. We then attempt verification using the temporary phone number previously used for registration from the receive-smss provider. When the attempt fails, we repeat the procedure using a personal phone number.

For \textit{credit-card–based} verification, the process's sensitivity limits our evaluation to testing whether a realistic but non-existent credit card number is accepted. 

\mypar{Results}
For \textit{email-based} verification, 3 out of 15 operators offer this mechanism. All three detect and reject the use of temporary email addresses. However, none of the operators allows us to successfully complete age verification using personal email accounts, indicating that the approach lacks the sensitivity required to meaningfully distinguish between eligible and ineligible users.
Only 1 out of 15 operators offers \textit{SMS-based} verification. This operator does not detect the use of a virtual number, allowing us to complete age verification without requiring demonstration of control over an adult-bound subscription. Moreover, the same phone number can be reused to obtain a new proof of age within a few minutes, including when initiating the flow from different countries.
For \textit{credit-card-based} verification,  5 out of 15 operators offer this mechanism. As all websites perform a transaction to validate the card, our test necessarily fails in all cases.

\mypar{Adversary relevance}
Temporary email addresses, virtual phone numbers, and shared payment instruments correspond to low-effort circumvention paths and are therefore realistic for motivated underage users (A1). Privacy-conscious adults (A2) may use email and SMS techniques to avoid disclosing more sensitive identity attributes. If accepted by providers, these signals are particularly attractive to AVaaS actors (A3), because they can support repeatable and low-cost creation of reusable verified accounts or proofs.

\section{Bypassing Verification}
\label{ssec:bypass}
We evaluate practical bypass strategies that (i) do not require defeating the underlying age-assurance provider cryptography or backend, and (ii) can be exercised by realistic adversaries in the wild. We focus on four bypass families: account reuse, session re-use, client-side gating that can be neutralised through browser tampering, and jurisdictional evasion via VPNs.

\subsection{Accounts and Verification Tokens} 

% =====================================================================
% (d) ACCOUNT / TOKEN SHARING  -- one verification, many users
% =====================================================================
\begin{figure}[h]
\centering
\begin{tikzpicture}
  % token (key icon) at the top center
  \node[font=\footnotesize, ok] at (4.5,1.85) {verified token / credentials};
  \tokenkey{(4.1,1.15)}{1}
 
  % three laptops in a horizontal row below
  \laptop{(1.5,-0.9)}{1.2}
  \laptop{(4.5,-0.9)}{1.2}
  \laptop{(7.5,-0.9)}{1.2}
  \node[font=\footnotesize, att] at (1.5,-1.2) {device 1};
  \node[font=\footnotesize, att] at (4.5,-1.2) {device 2};
  \node[font=\footnotesize, att] at (7.5,-1.2) {device N};
 
  % fan-out arrows from the token down to each laptop
  \draw[attflow] (4.25,0.7) -- (1.5,-0);
  \draw[attflow] (4.5,0.7) -- (4.5,-0);
  \draw[attflow] (4.75,0.7) -- (7.5,-0);
 
  % checkmarks beside each laptop
  \node[ok, tag] at (2.35,-0.7) {\checkmark};
  \node[ok, tag] at (5.35,-0.7) {\checkmark};
  \node[ok, tag] at (8.35,-0.7) {\checkmark};
 
\end{tikzpicture}
\caption{Overview of account attacks methodology.}
\label{sec:accountstokens}
\end{figure}
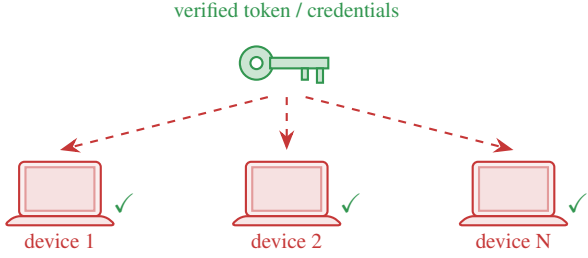

The first bypass strategy involves obtaining an account whose verification is handled by a third party. Here, our interest lies in studying shareability, e.g., the usage of the same account in parallel, and the circumvention of verification techniques.

\mypar{Methodology}
We distinguish between provider accounts and website accounts. Provider accounts are managed by providers and allow users to retain and re-present a previously issued proof of age. Website accounts are created directly on adult websites and may store an associated age-verified status, independently of the provider used to obtain it (See Fig.~\ref{sec:accountstokens}).

\noindent\textit{Provider accounts.}
From the set of providers considered in our study, we first identify those offering an account feature for storing proof of age. To assess shareability and concurrent reuse, we attempt to validate sessions on the same target site from multiple devices connected to different networks, using the same provider account.

\noindent\textit{Website accounts.}
We compile the 100 most visited sites by aggregating traffic across the four countries. We then identify sites that (i) support account creation and (ii) deploy at least one age-gating measure in at least one of the four countries. For each eligible site, we create an account using a temporary email address, verify age, and exclude sites without multimedia streaming to ensure comparable cross-device conditions.

We perform two analyses. (1) Whether the same website account can be used to stream content simultaneously from two devices connected to different networks. (2) For websites offering different age-verification options depending on the access country, whether an account verified in one country can be reused to stream content from another country where different verification methods are offered.

\mypar{Results}
Table~\ref{tab:av-provider-account} summarizes the provider-account characteristics. 
\noindent\textit{Provider accounts.}
8 out of 15 providers support the creation of an account that can retain and reuse proof of age. 
Not all are, however, compatible with our shareability experiment: \anonymage reject both the fake and our own IDs (supposedly as the French operator does not accept non-French IDs), not allowing us to create an account. \gotaca requires the mobile app to create an account, but the app does not yet fully integrate the procedure, preventing us from completing it.
Of the remaining 6, 2 are robust against trivial account sharing techniques, as \veratad allows age proof to be reused only through the use of a passkey, thereby raising the technical barrier, and \yoti asks to verify ownership of the account by performing and comparing a biometric facial scan (notably, this appears to imply retention of age verification data).
Of the remaining 4, 3 (\agego, \ageverif and \pleenk) permit account reuse across devices, while \verifymy relies on the \anonymage application for account creation and age-proof storage. For most operators, account access is gated by a one-time password sent to the registered email address or phone number at login time, with the exception of \pleenk.
Interestingly, 4 out of 8 providers (\agego, \ageverif, \anonymage and \pleenk) allow account creation using a temporary email address or temporary phone number.

\mypar{Adversary relevance}
Account reuse is most directly relevant to AVaaS actors (A3) as reusable provider or website accounts can be resold, shared across clients, or used as the basis for scalable access services. Motivated minors (A1) and privacy-conscious adults (A2) may also benefit from account reuse, but typically only when they can obtain access to an already verified account, which in practice often depends on sharing channels operated by A3-like actors.

\noindent\textit{Website accounts.}
28 of 100 websites both deploy age-gating and allow account creation. Of these, 27 permit registration with a temporary email address.
To support a controlled multi-device streaming experiment, we exclude 12 of these 27 sites: two because they are not primarily streaming services, and ten because they require age verification at every access (which effectively prevents sharing of verified accounts). Among the remaining 15 websites, 13 allow simultaneous streaming on different devices and networks using the same account. 
For cross-country reuse, 6 of 100 websites both allow account creation and vary age-verification measures by country. Of these, 3 allow an account verified in one country to be reused from another where that verification method is unavailable.

\begin{table}
\caption{Account-sharing results by provider requirements.}
\centering
\footnotesize
\resizebox{\columnwidth}{!}{
\begin{tabular}{l  c c c c}
\hline

\textbf{Provider} &
\multicolumn{3}{c}{\textbf{Requirements}}& \textbf{Allows sharing}\\

& \textbf{login verification} &
\textbf{ID and Selfie} &
\textbf{Temp. mail/phone} & \\
\hline
\ageverif    &    & & \checkmark & \checkmark\\
\agego       &   & & \checkmark & \checkmark\\
\gotaca      &   & \checkmark & & \na\textsuperscript{1}\\
\yoti        & \checkmark   & \checkmark & & \\
\verifymy    &   & \checkmark & \checkmark & \na\textsuperscript{1}\\
\anonymage   &   & \checkmark & \checkmark & \na\textsuperscript{1} \\
\veratad     & \checkmark(Passkey) & \checkmark & &\\
\pleenk      &   & & \checkmark & \checkmark\\
\hline
\multicolumn{5}{l}{\textsuperscript{1} No legitimate procedure allowed us to obtain a verified account to test.} \end{tabular}}
\label{tab:av-provider-account}
\end{table}

\subsection{Session re-use via cookie replay}

% =====================================================================
% (e) COOKIE REPLAY  -- session re-use across time / devices
% =====================================================================

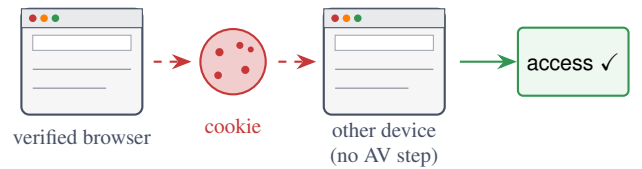
\begin{figure}[h]
\centering
\begin{tikzpicture} 
  % browser 1 -- verified session
  \browser{(0.8,0)}{0.8}{0.7}
  \node[font=\footnotesize, sys] at (0.8,-1.0) {verified browser};
 
  % cookie is the flowing artifact (no surrounding box)
  \cookieicon{(2.8,0)}{1.8}{att}
  \node[font=\footnotesize, att] at (2.8,-0.85) {cookie};

  % browser 2 -- replayed, no AV
  \browser{(4.8,0)}{0.8}{0.7}
  \node[font=\footnotesize, sys, align=center] at (4.8,-1.1)
       {other device\\(no AV step)};
 
  % accepted
  \node[okbox, minimum width=1.4cm, minimum height=0.9cm] (ok) at (7.3,0)
       {access \checkmark};
 
  \draw[attflow] (1.75,0) -- (2.25,0);
  \draw[attflow] (3.4,0)  -- (3.9,0);
  \draw[okflow]  (5.8,0) -- (ok);
\end{tikzpicture}
\caption{Overview of cookie replay attacks methodology.}
\label{fig:cookies}

\end{figure}
 
A second bypass class exploits persistent client state, i.e., session cookies, to remember that a device has already been age-verified. If a service issues a long-lived session token after verification, an adversary may reuse it or transfer it to another client, thereby rendering verification a shareable artifact. This enables verification sharing and supports automation via lightweight browser extensions or headless browser wrappers, which can distribute and refresh valid sessions.

\mypar{Methodology} We target websites that (i) deploy age-gating mechanisms and (ii) permit access without requiring a registered user account. After completing an age-gated access session, we extract and store browser cookies, then re-evaluate access by restoring the cookies after 24 hours and 7 days, respectively, from multiple devices and IP addresses in parallel (See Fig.~\ref{fig:cookies}). When necessary, we manually extend cookie expiration timestamps beyond the test time to assess for server-side validation.

\mypar{Results} Of the 100 assessed websites, 47 either do not offer content streaming or require an account to access, 26 do not implement age verification in any targeted jurisdiction, and 2 were unreachable at the time of testing. Of the remaining 26, 14 have expiration times under 24 hours, 1 has 7 days (allowing access the first time but not the second), and the remaining 11 have much longer expiration times, ranging from 100 days to 1 year. 10 of the 26 check for duplicated parallel sessions, while 16 do not. All enforce cookie expiration server-side.

\mypar{Adversary relevance}
Similarly to account reuse, cookie replay is less likely to be performed manually by most motivated underage users (A1), as it requires a valid cookie, but becomes realistic if packaged through browser extensions, shared profiles, or simple instructions by AVaaS actors (A3). Privacy-conscious adults (A2) with moderate technical ability can restore session state directly, however, they still require access to a valid, non-personal cookie (e.g., via A3).

\subsection{Client-side gating and page tampering}

% =====================================================================
% (f) CLIENT-SIDE GATING / PAGE TAMPERING  -- blur removed in the DOM
% =====================================================================
\begin{figure}[h]
    \centering
\begin{tikzpicture}

  % browser window 1 -- blurred content
  \begin{scope}[shift={(0.8,0)}]
    \draw[sysbox] (-1.1,-1.0) rectangle (1.1,1.0);
    \draw[fill=sys!10, draw=sys] (-1.1,0.7) rectangle (1.1,1.0);
    % traffic-light circles
    \fill[att]      (-0.95,0.85) circle (0.06);
    \fill[orange]   (-0.78,0.85) circle (0.06);
    \fill[ok]       (-0.61,0.85) circle (0.06);
    % blurred content area: noisy pattern
    \fill[pattern=north east lines, pattern color=sys!40]
          (-0.95,-0.85) rectangle (0.95,0.55);
    \node[sys, font=\footnotesize\bfseries] at (0,-0.15) {blurred};
  \end{scope}
 
  % "wrench" / DevTools edit
  \node[attbox, minimum width=1.7cm, minimum height=0.9cm] (edit) at (4.0,0)
       {\shortstack{remove\\overlay (DOM/CSS)}};
 
  % browser window 2 -- clear content
  \begin{scope}[shift={(7.0,0)}]
    \draw[okbox] (-1.1,-1.0) rectangle (1.1,1.0);
    \draw[fill=ok!15, draw=ok] (-1.1,0.7) rectangle (1.1,1.0);
    \fill[att]      (-0.95,0.85) circle (0.06);
    \fill[orange]   (-0.78,0.85) circle (0.06);
    \fill[ok]       (-0.61,0.85) circle (0.06);
    % clear content: simple "image" rectangles
    \draw[fill=ok!25, draw=ok] (-0.85,-0.15) rectangle (-0.05,0.5);
    \draw[fill=ok!25, draw=ok] (0.05,-0.15) rectangle (0.85,0.5);
    \draw[ok, thick] (-0.85,-0.45) -- (0.85,-0.45);
    \draw[ok, thick] (-0.85,-0.7)  -- (0.55,-0.7);
    \node[ok, font=\tiny\bfseries] at (0,0.85) {};
  \end{scope}
 
  \draw[attflow] (1.95,0) -- (edit);
  \draw[okflow]  (edit) -- (5.85,0);
\end{tikzpicture}
\vspace{0.01cm}
\caption{Overview of page tampering attacks methodology.}
\label{fig:blur}
\end{figure}
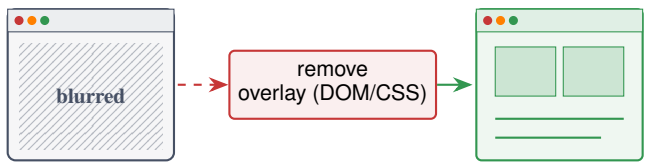

A third bypass family targets services that stream adult content to the end device and rely on client-side presentation controls (e.g., blur overlays) for gating. Here, users receive the sensitive asset regardless of their verification status, and restrictions are implemented only at the UI layer. Any architecture that transmits sensitive content before a server-side authorization decision assumes a benign client—an assumption incompatible with motivated evaders.

\mypar{Methodology} As above, we first compile a list of the 100 most visited sites by aggregating traffic across the four countries considered in this study. From this list, we remove websites that do not deploy age verification in any of the four countries or that do not provide content streaming.
We then classify whether a website delivers adult content to the client while applying only a blur prior to verification. For blur-based implementations, we evaluate whether it can be removed through straightforward page-level manipulation (See Fig.~\ref{fig:blur}).

\mypar{Results} Of 100 analyzed websites, 33 are not related to pornographic content streaming, and 26 deploy no age verification mechanism in any of the four countries. Of the remaining 41, 8 require a valid account; 17 either show only a static access template page (no content before verification) or provide only pre-blurred media assets. Among the remaining blur-based implementations, 8 deploy DOM-integrity mechanisms (e.g., observer-based reapplication of blur) to detect or revert page modifications; nonetheless, in 8 cases, we remove the blur via trivial client-side manipulations (e.g., deleting an overlay element or altering a CSS property), restoring clear playback.

\mypar{Adversary relevance}
Client-side page tampering, especially if guided, requires basic technical skills. Automation slightly raises the barrier. We therefore consider it qualitatively not accessible to underage users (A1), while more realistically available to privacy-conscious adults (A2). 
For AVaaS actors (A3), such bypasses are attractive because they can be packaged into lightweight scripts, extensions, or automated clients whenever the same site-side pattern recurs.

\subsection{Geofencing evasion via VPNs}
% =====================================================================
% (g) VPN GEOFENCING EVASION
% =====================================================================

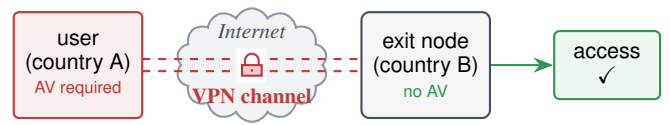
\begin{figure}[h]
    \centering
\begin{tikzpicture}

  \node[attbox, minimum width=1.7cm, minimum height=1.4cm] (cA) at (0.5,0)
       {\shortstack{user\\(country A)\\\textcolor{att}{\scriptsize AV required}}};
 
  % Internet cloud (behind the tunnel)
  \node[draw=sys!60, fill=sys!7, thick, cloud, cloud puffs=14,
        cloud puff arc=130, minimum width=2cm, minimum height=1.5cm,
        aspect=2.4] (net) at (2.8,0) {};
  \node[font=\footnotesize\itshape, sys] at (2.8,0.45) {Internet};
 
  % encrypted VPN channel (two parallel dashed lines through the cloud)
  \draw[att, thick, dashed] (1.35, 0.08) -- (5, 0.08);
  \draw[att, thick, dashed] (1.35,-0.08) -- (5,-0.08);
 
  % padlock in the middle of the channel
  \begin{scope}[shift={(2.8,0)}]
    \fill[white] (-0.20,-0.22) rectangle (0.20,0.22);
    \draw[att, thick] (-0.09, 0.0) -- (-0.09, 0.06) arc (180:0:0.09)
                      -- ( 0.09, 0.0);
    \draw[att, thick, fill=att!25] (-0.13,-0.14) rectangle (0.13, 0.0);
    \fill[att] (0,-0.07) circle (0.018);
  \end{scope}
  \node[font=\footnotesize\bfseries, att] at (2.8,-0.4) {VPN channel};
 
  % exit node country B
  \node[sysbox, minimum width=1.7cm, minimum height=1.4cm] (cB) at (5.1,0)
       {\shortstack{exit node\\(country B)\\\textcolor{ok}{\scriptsize no AV}}};
 
  % site
  \node[okbox, minimum width=1.4cm, minimum height=0.9cm] (site) at (7.5,0)
       {\shortstack{access\\ \checkmark}};
 
  \draw[okflow] (cB) -- (site);
\end{tikzpicture}
\vspace{0.01cm}
\caption{Overview of VPN attacks methodology.}
\label{fig:vpn}
\end{figure}

Virtual private networks (VPNs) are an effective one-step evasion strategy when age verification is enforced only by jurisdiction. By choosing an exit node in a country where the service does not mandate age verification or applies a less restrictive regime, users can often access the full service without completing any verification.

\mypar{Methodology}
We access each of the 154 websites from jurisdictions where age verification is not currently mandated using two independent commercial VPN providers to assess whether sites detect or block VPN traffic or, instead, enforce age verification regardless of apparent location (See Fig.~\ref{fig:vpn}).

\mypar{Results}
Among the 154 evaluated websites, only one appears to implement VPN detection, blocking both tested commercial VPN providers while allowing connections from cloud infrastructure used as VPN exit nodes. Two additional websites enforce age verification regardless of user location, applying verification uniformly across all tested ($>20$) jurisdictions.

\mypar{Adversary relevance}
VPN-based evasion is the most broadly accessible bypass in our threat model. It is realistic for motivated underage users (A1), who can use consumer VPNs or similar tools with little technical knowledge, and for privacy-conscious adults (A2), who may already rely on VPNs to reduce tracking or avoid identity disclosure. Notably, it may be considered by itself an AVaaS solution (A3). 

\section{Synthesis and Implications}
\label{sec:synthesis}

This section synthesizes the empirical results across verification mechanisms and deployment-level bypasses. We first map the evaluated approaches to the adversaries and security properties defined in Section~\ref{sec:threatmodel}. We then derive implementation red flags that explain why nominally compliant deployments fail in practice. Finally, we discuss the architectural role of double anonymity and why robustness and unlinkability must be considered jointly.

\begin{table}
\caption{Pentesting results and mapping with adversaries.}
\centering
\footnotesize
\resizebox{\linewidth}{!}{
\begin{tabular}{ccccc}
\toprule
                           & \textbf{Pentesting Results}    & \multicolumn{3}{c}{\textbf{Relevant Adversaries}}                                          \\
                              &                           & Motiv. Minor              & Priv.Consc. Adult         & AVaaS                     \\ \midrule
\textbf{Verif. Approaches}   &                \textbf{Vulnerable}         &                           &                           &                           \\ \midrule
Age Estimation             &\checkmark& \checkmark                    & \checkmark                    & \checkmark                    \\
ID Verification                 &\checkmark& \checkmark                    & \checkmark                    & \checkmark                    \\
Email                           & *                         & \checkmark & \checkmark & \checkmark \\
SMS                             &\checkmark& \checkmark & \checkmark & \checkmark \\
Credit Card                        &                           & \checkmark &                           &                           \\
Physical Ticket                 & ** &                           &                           & \checkmark \\
eID Login                    & **                        &                           &                           &                           \\ \midrule
\textbf{Bypass Approaches} &                   \textbf{Applicable}         &                           &                           &                           \\ \midrule
Account Sharing                   &\checkmark&                           &                           & \checkmark \\
Cookie Replay                       &\checkmark&                           &                           & \checkmark \\
Page Tampering                         &\checkmark&                           &               \checkmark            & \checkmark \\
VPN                                   &\checkmark& \checkmark                   & \checkmark                    &            \checkmark              \\ \bottomrule
\multicolumn{5}{l}{* The method never worked even in legitimate scenarios.} \\
\multicolumn{5}{l}{** The method could not be empirically assessed due to lack of availability.}

\end{tabular}}
\label{tab:resultstable}
\end{table}

\mypar{Security properties of deployed mechanisms}
As shown in Table~\ref{tab:resultstable}, most deployed age-verification mechanisms can be practically bypassed by a wide range of real-world adversaries, undermining their ability to prevent access to adult content under the threat model considered in this paper.

Table~\ref{tab:qualitative} summarizes how each verification approach performs with respect to the three properties defined in Section~\ref{sec:threatmodel}: robustness against evasion, resistance to replay or sharing, and unlinkability of identity and intent. We use a Low/Medium/High scale, where higher values indicate stronger security or privacy properties. For \textit{P1—Robustness}, Low means that the approach is exploitable in more than 75\% of providers, Medium that exploitation is observed in 25\%–75\% of cases, and High if exploitable in fewer than 25\%.
For \textit{P2—Anti-replay}, Low denotes methods that are trivially replicable, Medium indicates replication requiring technical expertise and effort, and High are assigned when no practical replay was identified.
Finally, for \textit{P3—Unlinkability}, Low indicates storage of directly identifying information, Medium potential indirect linkage, and High no digital storage of information.
\begin{table}
\caption{Qualitative security-property assessment of verification mechanisms. Higher values indicate stronger security or privacy properties under our threat model.}
\label{tab:qualitative}
\footnotesize
\centering
\resizebox{\columnwidth}{!}{
\begin{tabular}{lccc}
\toprule
\textbf{Verification approach} &
\textbf{P1: Robustness} &
\textbf{P2: Replay resistance} &
\textbf{P3: Unlinkability} \\
\midrule
Age Estimation              & Low    & Medium & Medium \\
ID Verification             & Medium & Medium & Low \\
Email\textsuperscript{1}    & N/A    & Low    & Medium \\
SMS                         & Low    & Low    & Medium \\
Credit Card                 & High   & High   & Low \\
Physical Ticket\textsuperscript{2} & N/A & Low & High \\
eID Login\textsuperscript{2}       & N/A & High & Low \\
\bottomrule
\multicolumn{4}{l}{\textsuperscript{1} The method did not work even in legitimate scenarios, so robustness is not assessed.} \\
\multicolumn{4}{l}{\textsuperscript{2} The method could not be empirically assessed due to lack of availability.}
\end{tabular}
}
\end{table}

Both \textit{age estimation} and \textit{ID verification} fail to fully satisfy the three properties. Age estimation provides low robustness because it is broadly vulnerable to camera-input manipulation, while ID verification provides only medium robustness because some providers rejected low-effort forgeries but others accepted fake or screen-recaptured documents. Both approaches also create replay or sharing risks when verification results can be persisted through accounts or tokens. From a privacy perspective, ID verification provides low unlinkability because it directly involves identifying documents, whereas age estimation provides only medium unlinkability because it relies on biometric processing even when no civil identity document is submitted.

\textit{SMS}-based verification provides low robustness and low replay resistance, since virtual numbers and number reuse make it easy to misuse. \textit{Email}-based verification could not be evaluated as a functioning protection mechanism because it did not work even in legitimate scenarios; however, if used as a reusable proof, it would still provide low replay resistance. \textit{Credit-card}-based verification was not bypassed in our experiments and provides stronger replay resistance, but it has low unlinkability because it links access to financial metadata and excludes users without suitable cards. 

\textit{Physical ticket}-based mechanisms could not be empirically validated due to the unavailability of purchase channels. Conceptually, they may provide high unlinkability if purchase and redemption remain separated, but low replay resistance because ticket use is decoupled from the person who passed the in-person check. \textit{Digital identity}-based verification was not observed in current deployments. It would likely provide strong robustness and replay resistance, but low unlinkability unless implemented through privacy-preserving attribute proofs or double-anonymity architectures.

\begin{answer}\textbf{Answer to RQ3:} With the exception of credit-card checks, all evaluated verification methods were susceptible to one or more attacks. Every provider exhibited at least one exploitable weakness, enabling each of the three identified adversary classes to bypass or exploit verification across all websites and in all jurisdictions, indicating that current deployments are ineffective against considered adversaries.
\end{answer}

\mypar{Implementation red flags} Regulatory instruments enumerate classes of age-assurance methods, but mechanism labels alone are a poor proxy for security and privacy outcomes under real implementations. Based on our empirical work, we outline red flags that strongly correlate with failures in robustness against evasion, anti-replay, and unlinkability. 

\mypara{Client-side-only gating (UI blur/overlay as enforcement)} If sensitive assets are delivered pre-verification and access control is enforced via a DOM/CSS layer, the design assumes a benign client and becomes easily bypassable through page-level tampering. In our tests, blur could be removed through simple manipulations in 8 cases. 

\mypara{Long-lived bearer artifacts after verification (cookie/session replay/tickets)}
  Any workflow that relies on persistent client state for verification, whether online or offline, risks turning verification into a transferable commodity. We observed that restoring previously issued cookies after $>$7 days still granted access in 11 out of 26 eligible cases, while in 16 cases, streaming was allowed from multiple devices with the same session cookie. Similarly, 13 out of 15 websites allow parallel streaming across the same account sessions.
  
\mypara{Account-based verification reuse and multi-device streaming}
  % Reusability reduces friction but often directly enables sharing: among tested sites supporting controlled multi-device experiments, 22 out of 26 allowed simultaneous streaming from different devices on different networks using the same account. Such designs naturally support A3-style resale of verified access. 
  Reusability reduces friction but often enables sharing: in controlled multi-device tests, 22/26 sites allowed simultaneous streaming from different devices and networks using the same account, supporting A3-style resale of verified access.

\mypara{Cross-country reuse without re-verification}
  We observed cross-site and cross-country reuse patterns, including cases where accounts verified under one regime remained valid when accessed from countries offering different verification measures, allowing users to pick the weakest verification method.

\mypara{Identity/document workflows that accept low-effort fakes or screen recapture}
  Providers that do not reject clearly fake documents (or accept a document photo re-captured from a screen) signal weak authenticity checks and limited resistance to scalable abuse. We observed operators accepting IDs readily obtainable online and recaptured from a screen.

\mypara{Easily automated enrollment signals (temporary email/phone acceptance)}
When websites/providers accept temporary contact channels for verification, adversaries can automate at scale and monetize verified accounts; we found widespread acceptance of disposable phone numbers as a stand-in for age verification in account creation, and a lack of validation of email addresses associated with accounts, allowing for the massive distribution of account credentials.

\mypara{“Double anonymity” claims without independently auditable data flows}
Even where double anonymity is advertised via a third-party provider, opaque implementations prevent users and auditors from verifying what data are shared and whether linkage is possible, failing to ensure absence of privacy exposure.

\begin{answer}
\textbf{Answer to RQ4:} 
The widespread lack of effective mitigations against bypass and redistribution indicates that adult-website age verification threat models do not align with real-world adversaries. Even with stronger verification methods, implementations often permit full circumvention (e.g., via VPNs or page tampering) and enable the anonymous sharing of credentials and bearer artifacts, making the redistribution of verified access trivial and low-effort.
\end{answer}

\mypar{Why stronger verification requires architectural separation (Double Anonymity)}
While our results demonstrate widespread deployment weaknesses, this outcome is unsurprising, as current age-verification systems largely act as imperfect surrogates for the class of mechanisms that could, in principle, satisfy the desired security properties—namely, robust digital identity frameworks such as SPID, BundID, FranceConnect, or One Login. However, while these infrastructures can theoretically support strong authentication, device binding, resistance to impersonation, and may further discourage account redistribution by tightly binding access to an individual’s real-world identity, they remain fundamentally ill-suited to the adult-content context because their use would directly or indirectly bind real-world identities to intimate online behavior, creating unacceptable privacy risks.

It is not our objective to advocate for a specific solution. However, we conclude that the fundamental weakness lies less in individual verification techniques than in the prevailing system architecture. Current designs tightly couple identity validation and website access within a single opaque workflow, forcing trade-offs whereby improvements in one dimension (e.g., accuracy or resistance to bypass) degrade others (e.g., non-linkability).
By contrast, transparent double-anonymity architectures—where identity validation and content access are handled by distinct, non-colluding entities—can, absent implementation flaws, theoretically satisfy all the identified security and privacy requirements. In such designs, strong identity checks can be performed without revealing the target website, while content providers receive only unlinkable age assertions rather than personal data. This separation enables the use of stronger verification mechanisms, eliminating entire classes of bypass attacks rooted in weak verification, while also substantially limiting the impact of breaches, as no single actor can reconstruct both user identity and browsing behavior. 
Crucially, this architectural separation also supports auditability: unlinkability claims can be evaluated by inspecting data flows between providers. In contrast, most of today’s deployments resist scrutiny, as assertions of data minimization, immediate deletion, or double anonymity are not readily auditable by users and regulators\footnote{Note that one provider offers the open source code of its implementation in a public repository, partially providing auditability.}. Notably, as shown in Table~\ref{tab:regulatory-methods}, France's ARCOM and Italy's AGCOM are the only instruments requiring websites to provide at least one double anonymity method.

\section{Threats to Validity}
\label{sec:ttv}

Like any empirical study of deployed mechanisms, this work has limitations that may influence its results:

\mypar{External Validity} Our empirical evaluation is based on a finite and evolving set of adult websites and age verification deployments observed during a specific time window. Age verification ecosystems are highly dynamic: websites frequently change providers, adjust enforcement policies, or deploy new countermeasures in response to regulatory pressure or public scrutiny. Our sample is also derived from Semrush traffic estimates and adult-category classifications, which may undercount or misclassify adult websites, especially in the presence of mirrors, regional domains, redirects, or anti-measurement practices. Nevertheless, our contribution concerns recurring architectural and design choices. We expect the high-level insights to remain valid as systems evolve.

\mypar{Internal Validity} Our methodology relies on black-box testing. Some observed behaviors may be influenced by undisclosed factors. While this uncertainty may affect the precise attribution of causes, it does not undermine that access was granted under conditions that should have triggered denial.

\mypar{Measurement Bias}
Some aspects of our measurements depend on client-side observations, such as cookie persistence, browser state, and access decisions. Unidentified variables could influence results. To mitigate this risk, we repeated tests across multiple contexts and focused on consistent behaviors. However, we cannot rule out the possibility that some edge cases were overlooked or disproportionately represented.

\section{Conclusion}
\label{sec:conclusion}

Age verification has rapidly become a cornerstone of regulations aimed at limiting minors’ access to online pornographic content. Yet, as this work shows, the effectiveness of today’s deployed systems is more often presumed than proven. Through a systematic, multi-country evaluation of adult websites and age verification providers, we find that these assumptions do not hold in practice.
Across jurisdictions, compliance is inconsistent and techniques are heterogeneous. Most critically, no evaluated mechanism reliably withstands realistic adversaries, enabling evasion and redistribution of verified access. Moreover, even assuming strong verification in isolation, website integrations frequently reintroduce trivial bypasses. Beyond their limited protective value, these systems impose substantial privacy risks, requiring users to accept exposure for assurances that our empirical analysis finds fragile at best.

Crucially, our results indicate that these shortcomings are not due to implementation choices. Current designs tightly couple identity validation and content access, forcing trade-offs between robustness and unlinkability that architecturally favor superficial compliance over meaningful protection. 
Taken together, our results suggest that regulation-mandated age verification for adult websites currently functions as compliance theater, failing to effectively protect minors while elevating privacy risk, offering security assurances that do not withstand empirical scrutiny.

\section*{Acknowledgments}
The authors would like to thank Daniel Ponticello for the insightful discussion that eventually led to the development of this work.
%-------------------------------------------------------------------------------
% \section*{LLM Usage Considerations}
% Large Language Models (LLMs) were used solely for text validation and minor language refinement. The authors are fully aware of the potential risks associated with using LLMs, including issues related to accuracy and bias, and take full responsibility for the content of this paper. No part of the scientific reasoning, data analysis, or experimental results was generated or influenced by LLMs.

%-------------------------------------------------------------------------------

%-------------------------------------------------------------------------------

\bibliographystyle{plainurl}
\bibliography{bib}

% optional clearing of the page
\appendix

% \newpage
\section*{Ethical Considerations}
\label{sec:ethics}

This paper adopts an offensive security perspective to evaluate deployed age-verification systems on adult websites. Because these systems aim to protect minors while processing highly sensitive personal data, we conducted the study using a stakeholder-based ethics analysis aligned with the Menlo Report principles.

\subsubsection*{Stakeholders and Impacts}

\mypar{Minors}
\emph{During:} Our tests did not involve minors, underage accounts, or third-party credentials.
\emph{After:} Actionable bypass details could lower circumvention costs, but documenting empirical weaknesses can also prevent misplaced trust in fragile controls and motivate stronger protections.

\mypar{Adult end users subject to age verification}
\emph{During:} We did not collect, access, or attempt to deanonymize other users’ data. Where flows required selfies or videos, we used only consenting adult researchers’ own inputs and no leaked, scraped, or third-party personal data.
\emph{After:} Evidence of bypassable deployments may reduce reliance on systems that impose privacy costs without corresponding protective benefits.

\mypar{Website operators and age-verification providers}
\emph{During:} We limited interactions to controlled tests needed to validate security properties, avoiding denial-of-service, account compromise, large-scale automation, and access to third-party accounts or data.
\emph{After:} Providers and sites may face reputational harm, but also gain actionable evidence to prioritize mitigations. Relevant providers and sites were contacted before submission to enable remediation.

\mypar{Regulators and policymakers}
\emph{During:} No direct impact identified.
\emph{After:} Findings may inform enforcement and procurement standards, but could also be misused to justify either more intrusive identity binding or premature dismissal of privacy-preserving approaches. We therefore frame results as a security-and-privacy measurement contribution rather than a specific policy prescription.

\subsubsection*{Ethical harms considered}

\mypar{Beneficence}
The main risk is lowering the barrier to evasion or enabling AVaaS-style commercialization; secondary risks include remediation costs and targeted abuse against specific deployments.

\mypara{Mitigations}
We do not release exploit tooling, reusable artifacts, or step-by-step instructions. We report outcomes, attack classes, and architectural red flags at a level intended to support remediation without serving as a bypass guide. We also conducted coordinated disclosure and documented acknowledgements\footnote{At the time of publication, 12 out of 37 contacted entities acknowledged the disclosure and indicated that they are taking or planning remedial actions. Of these, 2 providers and 1 website actively engaged with us to discuss improvements.}.

\mypar{Respect for Persons}
Age-verification workflows involve sensitive data and may create privacy intrusion, unwanted linkage between identity and lawful behavior, and chilling effects. Offensive testing also risks retaining sensitive artifacts or encouraging unsafe replication.

\mypara{Mitigations}
We used only consenting adult researchers’ inputs, avoided non-consensual personal data, and did not identify, track, or deanonymize users. We retained only evidence needed to validate findings and support aggregate reporting, and we avoided publishing session identifiers, reusable cookies/tokens, or other replay-enabling artifacts.

\mypar{Justice}
Age-verification failures impose asymmetric burdens: ineffective gates reduce protection for minors, while intrusive or brittle gates burden adults who are privacy-vulnerable or lack accepted documents, devices, or payment instruments. Measurement bias could also distort policy conclusions.

\mypara{Mitigations}
We report results at the level of attack classes and deployment patterns, using cross-site and cross-provider evidence rather than isolated edge cases. We interpret findings as evidence about deployments and threat models, not as claims about any user group’s behavior or intent.

\mypar{Respect for Law and Public Interest}
The study interacted with live systems in a sensitive regulatory space, creating risks around contractual violations, excessive automation, or framing results as an evasion blueprint.

\mypara{Mitigations}
We limited testing to controlled, non-disruptive interactions, avoided account compromise and denial-of-service, and did not conduct large-scale or persistent abuse. We frame the work in the public interest: improving regulatory fidelity, enabling remediation, and supporting child-protection goals without unnecessary privacy intrusion.

\mypar{Residual risk}
Despite these safeguards, public descriptions of bypass classes may still be adapted by determined adversaries, and we cannot fully prevent misuse or unsafe third-party replication.

\subsubsection*{Decision to proceed and to publish}

We considered “do not do” and “do not publish” alternatives. We proceeded because age verification is being deployed at scale, and measuring whether it satisfies its stated security properties is necessary to prevent compliance theater, resist normalization of intrusive identity or biometric collection without protective benefit, and support more equitable system requirements.

Accordingly, we judge publication ethically justified when paired with coordinated disclosure and minimization of operational exploitability.
% optional clearing of the page
%\input{sections/appendix}

\section*{Open Science}
\label{sec:openscience}
To support follow-up research and to provide further details on the results of our experiments, we release the dataset with the per-website and per-provider outcomes of each of our assessments at the following link \url{https://doi.org/10.5281/zenodo.20183951}. 
The public repository contains an .xls file with the \textit{redacted} version of the dataset, to limit misuse of our results. A \textit{non-redacted} version of the dataset is available at \url{https://doi.org/10.5281/zenodo.20184277}, accessible after verifying that the requester’s intent is consistent with research or auditing.

%%%%%%%%%%%%%%%%%%%%%%%%%%%%%%%%%%%%%%%%%%%%%%%%%%%%%%%%%%%%%%%%%%%%%%%%%%%%%%%%
\end{document}